\documentclass[journal,12pt,onecolumn,draftclsnofoot]{IEEEtran}
\IEEEoverridecommandlockouts
\usepackage[cmex10]{amsmath}
\usepackage{amsfonts}
\usepackage{amssymb}
\usepackage{amscd}
\usepackage{graphicx}
\usepackage{epsfig}
\usepackage{mathrsfs}
\usepackage{afterpage}
\usepackage{cite}
\usepackage{enumerate}
\usepackage{url}
\usepackage{mathtools}
\let\labelindent\relax
\usepackage{enumitem}
\usepackage{footnote}

\newcommand{\nn}{\nonumber\\}

\newcommand{\Y}{{Y}}
\newcommand{\Z}{{Z}}

\newcommand{\E}{\mathbb{E}}

\newcommand{\EE}{\mathcal{E}}
\newcommand{\A}{\mathcal{A}}
\newcommand{\N}{{N}}
\newcommand{\F}{\mathcal{F}}
\newcommand{\tF}{\tilde{\mathcal{F}}}
\newcommand{\Fk}{\mathfrak{F}}
\newcommand{\G}{\mathcal{G}}

\newcommand{\tin}{t\in [0,T]}
\newcommand{\de}{\triangleq}
\newcommand{\1}{\mathbf{1}}
\newcommand{\aeq}{\mathbf{\overset{(a)}{=}}}

\newcommand{\iin}{i\in \{1,2\}}
\newcommand{\kin}{k\in \{1,2,3,4\}}

\newtheorem{Theorem}{Theorem}
\newtheorem{Remark}{Remark}
\newtheorem{Lemma}{Lemma}
\newtheorem{Def}{Definition}
\newtheorem{Corollary}{Corollary}

\def\eor{\hfill\ensuremath{\Diamond}}

\begin{document}
\title{Functional Covering of Point Processes}
\author{\IEEEauthorblockN{Nirmal V. Shende} and
\IEEEauthorblockN{Aaron B. Wagner}
\thanks{N. V. Shende is with Marvell Technology, Inc., Santa Clara, CA 95054 (email: nshende@marvell.com). A. B. Wagner is with the School of Electrical and Computer Engineering, Cornell University, Ithaca, NY 14853
(email:wagner@cornell.edu). This work was performed when N. V. Shende was a student
at Cornell University. This paper was presented in part at the IEEE International Symposium on Information Theory, Paris, Jul. 2019~\cite{Shende19}. This research was supported by the US National Science Foundation under grants
CCF-1956192, CCF-2008266, and CCF-1934985.}}
\maketitle
\begin{abstract}
We introduce a new distortion measure for point processes called functional-covering distortion. It is inspired by intensity theory and is related to both the covering of point processes and logarithmic-loss distortion.   We obtain the distortion-rate function with feedforward under this distortion measure for a large class of point processes. For Poisson processes, the rate-distortion function is obtained under a general condition called constrained functional-covering distortion, of which both covering and functional-covering are special cases. Also for Poisson processes,  
we characterize the rate-distortion region for a two-encoder CEO problem and show that feedforward does not enlarge this region. 
\end{abstract}
\section{Introduction}

The classical theory of compression~\cite{Berger:RD} focuses on discrete-time,
sequential sources. The theory is thus well-suited to text, audio, 
speech, genomic data, and the like. Continuous-time signals are typically
handled by reducing to discrete-time via projection onto a countable
basis. Multi-dimensional extensions enable application to
images and video.

Point processes model a distinct data type that appears in diverse
domains such as neuroscience~\cite{johnson1996point, goldwyn2012point,
farkhooi2009serial,Sarma10,brown2004multiple,Rieke:Spikes}, 
communication networks~\cite{Giles:BTQ,Shahzad:Flow,Zhu:Flow},
imaging~\cite{Borcs:Point,Yu:Point}, blockchains~\cite{Nakamoto:Bitcoin,
Lewenberg:Poisson, Kawase:Poisson, Decker:Poisson}, and 
photonics~\cite{Laourine:Poisson:ISIT,
Wyner:Poisson:I,
Wyner:Poisson:II,
Lapidoth:Poisson:Feedback,
Shende:Poisson:Multiple}.
Formally, a point process can be viewed as a random counting measure
on some space of interest~\cite{Baccelli:Palm:Stat}, or if the space is a 
real line, a random counting function; we shall adopt the latter
view. Informally, it may be viewed as 
simply a random collection of points representing epochs in time or
points in space.

Compression of point processes emerges naturally in several of the
above domains. Sub-cranial implants need to communicate the 
timing of neural firings to a monitoring station over a wireless
link that is low-rate because it must traverse the 
skull~\cite{Sutardja:Cranial:JSSC,Skrivervik:Cranial}.
In network flow correlation analysis, one cross-correlates packet timings 
from different links in the network~\cite{Zhu:Flow}; 
this requires communication
of the packet timings from one place to another.
Compressing point process realizations in 2-D (also known
as \emph{point clouds}) arises in computer 
vision~\cite{deQueiroz:PointCloud,Golla:PointCloud,Tu:PointCloud}, 
and so on.

Various specialized approaches have been developed
for compressing point processes, and in particular
for measuring distortion. One natural approach is
for the compressed representation to be itself a
point-process realization. In this case, the distortion can be 
the sum of the absolute value of the differences
between the actual and  reconstructed epochs, with the constraint that the two processes
must have the same number of points.  For the Poisson point process, Gallager~\cite{Gallager76} obtained a lower bound on the rate-distortion function by insisting on the causal reconstruction of the points but allowing for their reorder. Bedekar~\cite{Bedekar'01} determined the rate-distortion function with the additional constraint of exact orders of epochs in reconstruction. Verd\'u~\cite{Verdu'96} allowed the reconstruction to be non-causal.   Coleman \emph{et al.}~\cite{Coleman'08} introduced the queueing distortion function, where the reproduced epochs lead the actual epochs. Rubin~\cite{Rubin1974} used the $L_1$ distance between the counting functions as a distortion measure. In a more general setting, Koliander \emph{et al.}~\cite{Koliander2018} gave upper and lower bounds on the rate-distortion function  under a more generic distortion defined between pair of  point processes. 

Most relevant to the present paper,
Lapidoth \emph{et al.}~\cite{Lapidoth2015} introduced a \emph{covering} distortion measure, where the reconstruction of a point process on $[0,T]$ is a subset $Y$ of $[0,T]$ that must contain all the points, and the distortion is the Lebesgue  measure of the covering set (see also Shen \emph{et al.}~\cite{Shen22}).

If we encode the subset $Y$ as an indicator function
\begin{equation*}
    Y_t = \begin{cases}
        1 & \text{if $t \in Y$} \\
        0 & \text{otherwise,}
    \end{cases}
\end{equation*}
then $Y_t = 0$ guarantees that no point occurred at time $t$, while $Y_t = 1$ indicates
that a point may occur at $t$. More generally, $Y_t$ could encode the relative belief
that there is a point at $t$. Inspired by this observation, and the notion of 
logarithmic-loss distortion~\cite{Courtade2011,Courtade2014}, we consider the following formulation.
For a realization of a counting (or point) process $y_0^T=(y_t:\tin)$ (i.e., $y_t$ is integer-valued, non-decreasing, and has unit jumps) and a non-negative reconstruction $\hat{y}_0^T$, we  define the \emph{functional-covering distortion} as
\begin{align}
d(\hat{y}_0^T,y_0^T)\de\int_0^T \hat{y}_t\,dt-\log(\hat{y}_t)\,dy_t.  
\label{eq:the_dist}
\end{align}
This is related to the covering distortion measure in the following sense. If we impose that $\hat{y}_t\in\{0,1\}$, then (\ref{eq:the_dist}) reduces to the covering distortion measure. Yet it is natural to consider the distortion in~(\ref{eq:the_dist}) without such a restriction, or with a more general set of allowable values for $\hat{y}_t$. In fact,
there are advantages to not restricting $\hat{y}_0^T$ to the set $\{0,1\}$. Consider a remote source setting where the encoder cannot access the point-process source directly, but instead observes a thinned version  where some of the points in the source point process are deleted randomly. Then, in case of the covering distortion the reconstruction can only be the entire interval $[0,T]$ (i.e. $\hat{y}_t=1,\tin$). On the other hand, under functional covering distortion the problem has a nontrivial solution.

The relation functional covering distortion measure to logarithmic-loss is as follows. If we constrain $\hat{y}_0^T$ to be bounded, then we can use a Girsanov-type transformation~\cite[Chapter VI, Theorems T2-T4]{Bremaud} to define a probability measure on the set of all counting processes using $\hat{y}_0^T$, and the distortion can be defined as the expectation of the negative logarithm of the Radon-Nikodym derivative between this probability measure and an appropriately chosen reference measure,
evaluated at the source realization, which is equivalent to (\ref{eq:the_dist}).
However,  we will allow $\hat{y}_0^T$ to be unbounded but integrable $\E[\int_0^T \hat{Y}_t\,dt]<\infty$. 

The relation to intensity theory is as follows.
Heuristically, given a random variable $M$, the \emph{intensity} of a point process represented by a counting function $Y_0^T$ is a non-negative process $\Gamma_0^T$ such that
$P( Y_{t+\Delta}-Y_t=1|M,Y_0^t)\approx \Gamma_t\Delta$ (see~Definition~\ref{def:int} for the precise statement). From~(\ref{eq:the_dist}), we expect any optimal $\hat{Y}_0^T$ (in the rate-distortion trade-off sense) to be related to the intensity of $Y_0^T$. In fact, we will see in the proof of Theorem~\ref{thm:R-D} that an optimal reconstruction $\hat{Y}_0^T$ is the intensity of $Y_0^T$ given the encoder's output. 

%

Beyond the introduction of the functional covering distortion measure and the accompanying
coding theorems, the paper provides a collection of results for the information-theoretic
analysis of point processes, which may be of independent use.
One such contribution is Theorem~\ref{thm:MI},  where we derive the mutual information between   point-processes with intensities and arbitrary random variables. This is the most general expression available for mutual informations involving point process with intensities. Theorem~\ref{thm:MI} subsumes the existing formulae for mutual informations  involving doubly stochastic Poisson processes~\cite{Kabanov,Davis,Shende-Wagner'17} and queuing processes~\cite{Sundaresan} as special cases. The other theorems proved in this paper are: we obtain the rate-distortion trade-off with feedforward for the functional-covering distortion measure  for  point processes which admit intensities (see Theorem~\ref{thm:R-D}). For Poisson processes, we obtain the rate-distortion region when the reconstruction function $\hat{y}_0^T$ is constrained to take value in a subset of reals (Theorem~\ref{thm:Poisson_Cons}). The covering distortion in~\cite[Theorem~1]{Lapidoth2015} is a special case of this constrained functional-covering distortion, hence the rate-distortion function in~\cite{Lapidoth2015} can be obtained as the special case of this theorem.  
We characterize the rate-distortion region for a two-encoder Poisson CEO problem (see Figure~\ref{fig:Poisson_CEO}) under functional-covering distortion  in Theorem~\ref{thm:Poisson_CEO}. To prove the converse of the CEO problem, we derive a strong data processing inequality for Poisson processes under superposition (see Theorem~\ref{thm:add_int}), which complements the strong data processing inequality for Poisson processes under thinning due to Wang~\cite{Wang'17}. We also provide a self-contained proof of Wang's theorem in Theorem~\ref{thm:thin_int}. The solution to the CEO problem gives   the rate-distortion trade-off for remote Poisson sources as an immediate corollary.

\section{Preliminaries}
We will consider  a probability space $(\Omega,\F,P)$   on which all stochastic processes considered here are defined. For a finite $T>0$, let $(\mathcal{F}_t:\tin)$ be an increasing family of $\sigma$-fields with $\F_T\in\F$.   We will assume that the given  filtration $(\F_t:\tin)$, $P$,  and $\F$  satisfy the ``usual conditions"\cite[Chapter III, p. 75]{Bremaud}: $\F$ is  complete with respect to $P$, $\F_t$ is right continuous, and $\F_0$ contains all the $P$-null sets of $\F_t$. Stochastic processes are denoted as $\hat{Y}_0^T=\{\hat{Y}_t:0 \le t \le T\}$. 
The process $X_0^T$ is said to be \emph{adapted} to the  history $(\mathcal{F}_t:t\in [0,T])$ if $X_t$ is $\F_t$ measurable for all $t\in[0,T]$.
 The internal history recorded by the process $X_0^T$ is denoted by $\mathcal{F}^X_t=(\sigma(X_s): s \in [0,t])$, where $\sigma(A)$ denotes the  $\sigma$-field generated by  $A$.  

A process $X_0^T$ is called $(\F_t:t\in [0,T])$-\emph{predictable} if $X_0$ is $\F_0$ measurable and the mapping $(t,\omega)\to X_t(\omega)$ defined from $(0,T)\times\Omega$ into $\mathbb{R}$ (the set of real numbers) is measurable with respect to the $\sigma$-field over $(0,T)\times\Omega$ generated by rectangles of the form 
\begin{align}
(s,t]\times A; \quad 0< s \leq t \leq T, \quad A \in \F_s.
\label{EQ:Pred_field}
\end{align}
For two measurable spaces $(\Omega_1,\F_1)$ and $(\Omega_2,\F_2)$, the product space is denoted by $(\Omega_1 \times \Omega_2,\F_1\otimes\F_2)$. We say that $A \rightleftarrows B \rightleftarrows C$ forms a Markov chain under measure $P$ if $A$ and $C$ are conditionally independent given $B$ under $P$. $P\ll Q$ denotes that the probability measure $P$ is absolutely continuous with respect to the measure $Q$. $\textbf{1}\{\mathsf{E}\}$ denotes the indicator function for an event $\mathsf{E}$.  $\log(x)$ is the natural logarithm of $x$. $(x)^+$ and $(x)^-$ denote the positive ($\max(x,0)$) and the negative part ($-\min(x,0)$) of $x$ respectively.  $\lceil x \rceil$ denotes the ceiling of $x$.  Throughout this paper we will adopt the convention that $0\log(0)=0$, $\exp(\log(0))=0$, and $0^0=1$. 

\begin{Def}
$\phi(x)=x\log(x)$ with convention that $0\log(0)=0$.
\end{Def}
We note that $\phi(x)$ is convex.

We will  use the following form of Jensen's inequality~\cite[Theorem 7.9, p. 149]{Klenke} and~\cite[Theorem 8.20, p. 177]{Klenke}.
\begin{Lemma}
If $f(x)$ is a convex function and $\E|X|<\infty$  then  $\E[f(X)]$ exists and for any two $\sigma$-fields $A$ and $B$,
\begin{align*}
\E[f(X)]\ge\E[f(\E[X|A,B])]\ge\E[f(\E[X|A])]\ge f(\E[X]).
\end{align*}
\end{Lemma}
We now recall the definition of mutual information for general ensembles and
its properties.
Let $A$, $B$, and $C$ be measurable mappings defined on a given probability space $(\Omega,\F,P)$, taking values in $(\mathcal{A},\mathfrak{F}^A)$, $(\mathcal{B},\mathfrak{F}^B)$, and $(\mathcal{C},\mathfrak{F}^C)$ respectively.  
Consider partitions of $\Omega$, $\mathfrak{Q}_A=\left\{\mathtt{A}_i, 1\le i \le N_A\right\}\subseteq\sigma(A)$ and $\mathfrak{Q}_B=\left\{\mathtt{B}_j, 1\le j \le N_B\right\}\subseteq\sigma(B)$. Wyner defined the conditional mutual information $I(A;B|C)$  as~\cite{WYNER197851} 
\begin{align}
I(A;B|C)=\sup_{\mathfrak{Q}_A,\mathfrak{Q}_B} \E\left[\sum_{i,j=1,1}^{N_A,N_B}P(\mathtt{A}_i,\mathtt{B}_j|C)\log\left(\frac{P(\mathtt{A}_i,\mathtt{B}_j|C)}{P(\mathtt{A}_i|C)P(\mathtt{B}_j|C)}\right)\right],
\label{EQ:MI_Def}
\end{align}
where the supremum is over all such partitions of $\Omega$. 
Wyner  showed that $I(A;B|C)\ge 0$  with equality if and only if $A \rightleftarrows C \rightleftarrows
 B $ forms a Markov chain \cite[Lemma 3.1]{WYNER197851}, 
and that  (what is generally referred to as) Kolmogrov's formula holds \cite[Lemma 3.2]{WYNER197851}
\begin{align}
I(A,C;B)=I(A;B)+I(C;B|A).
\label{EQ:Kolmogrov}
\end{align}
Hence if $I(A;B)<\infty$, then $I(C;B|A)=I(A,C;B)-I(A;B)$. The data processing inequality can be obtained from (\ref{EQ:Kolmogrov}) as well: if $A \rightleftarrows C \rightleftarrows
 B $ forms a Markov chain, then $I(A;B)\le I(C;B)$.

Denote by $P^{A,B}$, the joint distribution of $A$ and $B$ on the space ($\mathcal{A}\times \mathcal{B},\Fk^A\otimes\Fk^B$ ),  i.e.,
\begin{align*}
P^{A,B}(dA\times dB)=P((A^{-1}(dA),B^{-1}(dB)), \quad dA\in \Fk^A, dB\in\Fk^B.
\end{align*} 
Similarly, $P^A$ and $P^B$ denote the marginal distributions.
 Gelfand and Yaglom \cite{Gelfand} proved that   if $P^{A,B}\ll P^A\times P^B$, then the mutual information $I(A;B)$ (defined via (\ref{EQ:MI_Def}) by taking $\sigma(C)$ to be the trivial $\sigma$-field) can be computed as:
\begin{align}
I(A;B)=\E\left[\log\left(\frac{dP^{A,B}}{d(P^A\times P^B)}\right)\right].
\label{EQ:MI_LLR}
\end{align}
A sufficient condition for $P^{A,B}\ll P^A\times P^B$ is that $I(A;B)<\infty$ \cite[Lemma 5.2.3, p. 92]{Gray}.
We will also require the following result \cite[Lemma 2.1]{WYNER197851}:
\begin{Lemma}[Wyner's Lemma]
\label{Le:Wyner}
If $M$ is a finite alphabet random variable, then
\begin{align*}
I(M;U_0^T)=H(M)-\E\left[H(M|U_0^T)\right],
\end{align*}
where 
\begin{align*}
H(M|U_0^T)=-\sum_{m}P(M=m|U_0^T)\log\left(P(M=m|U_0^T)\right),
\end{align*}
\end{Lemma}
 and $H(M)$ is the entropy of $M$.

\section{Point Processes, Intensities, and Mutual Information}
Let $\mathcal{N}_0^T$ denote the set of counting realizations (or point-process realizations) on $[0,T]$, i.e., if ${N}^T_0\in\mathcal{N}_0^T$, then for $t\in[0, T]$, ${N}_t\in \mathbf{N}$ (the set of non-negative integers), is right continuous, and has unit increasing jumps with ${N}_0=0$. 
Let $\Fk^N$ be the restriction of the $\sigma$-field generated by the Skorohod topology on $D[0, 1]$ to $\mathcal{N}_0^{T}$. 
\begin{Def}
\label{def:int}
If $N_0^T$ is a counting process adapted to the history $(\F_t:\tin)$, then $N_0^T$ is said to have  $(P,\F_t:\tin)$-\emph{intensity} ${\Gamma}_0^T=(\Gamma_t:\tin)$, where $\Gamma_0^T$ is a non-negative measurable process if
\begin{itemize}
\item $\Gamma_0^T$ is $(\F_t:t\in [0,T])$-predictable,
\item $\int_0^T\Gamma_t\,dt < \infty$, $P$-a.s.,
\item and for all non-negative $(\F_t:\tin)$-predictable processes $C_0^T$:\footnote{The limits of the Lebesgue-Stieltjes integral $\int_a^b$ should be interpreted as $\int_{(a, b]}$.}
\begin{align*}
\E\left[\int_0^T C_s \, d\N_{s}\right]=\E\left[\int_0^T C_s \Gamma_s \, ds\right]. 
\end{align*}
\end{itemize}
\end{Def}
When it is clear from the context, we will drop the probability measure $P$ from the notation and  say $N_0^T$ has  $(\F_t:\tin)$-intensity $\Gamma_0^T$. 
\begin{Def}
A point process $Y_0^T$ is said to be \emph{Poisson process with rate $\lambda$} if its $(\F_t^Y:\tin)$-{intensity}  is $(\lambda:\tin)$.
\end{Def}
The above definition can be shown to imply  the usual definition of Poisson process~\cite[Theorem T4, Chapter II, p. 25]{Bremaud} and vice versa~\cite[Section 2, Chapter II, p. 23]{Bremaud}.

\begin{Def}
$P_0^{Y_0^T}$ denotes the distribution of  a point process $Y_0^T$ (on the space $(\mathcal{N}_0^T,\Fk^N)$) under which $Y_0^T$ is a Poisson process with unit rate.
\end{Def}
A point processes with stochastic intensity  and a Poisson process with unit rate are linked via the following result.  
\begin{Lemma}
\label{lem:abs_cont}
Let $P^{Y_0^T}$ be the distribution of a  point process $Y_0^T$ such that $P^{Y_0^T} \ll P_0^{Y_0^T}$. Then there exists a non-negative predictable process $\Lambda_0^T$ such that
\begin{align*}
\frac{dP^{Y_0^T}}{dP_0^{Y_0^T}}=\exp\left(\int_0^T\log(\Lambda_t)\,dY_t-\Lambda_t+1\,dt\right).
\end{align*}
Moreover, the $(P^{Y_0^T},(\F_t^Y:\tin))$-intensity of $Y_0^T$ is $\Lambda_0^T$. 
Conversely, if the $(P^{Y_0^T},\F_t^Y:\tin)$-intensity of $Y_0^T$ is $\Gamma_0^T$ and $\E_{P^{Y_0^T}}[\int_{0}^T|\phi(\Gamma_t)|\,dt]<\infty$, then $P^{Y_0^T} \ll P_0^{Y_0^T}$, and the corresponding Radon-Nikodym derivative is given by the above expression, where 
\begin{align*}
\E_{P^{Y_0^T}}\left[\int_0^T|\Gamma_t-\Lambda_t|\,dt\right]=0,\quad \E_{P^{Y_0^T}}\left[\int_0^T\1\{\Gamma_t\ne\Lambda_t\}\,dY_t\right]=0.
\end{align*}
In the latter case,
\begin{align*}
\E_{P^{Y_0^T}}\left[\log\left(\frac{dP^{Y_0^T}}{dP_0^{Y_0^T}} \right) \right]=\E_{P^{Y_0^T}}\left[\int_0^T\phi(\Gamma_t)-\Gamma_t+1\,dt\right].
\end{align*}
\end{Lemma}
\begin{IEEEproof}
Please see the supplementary material.
\end{IEEEproof}
 
The following theorem allows us to express the mutual information involving a point processes with intensity and other random variables in terms of the intensity functions. The proof of the theorem is similar to the proof of Theorem 1 in~\cite{Shende-Wagner'17}.

\begin{Theorem}
\label{thm:MI}
Let $Y_0^T$ be a point process with $(\F_t^Y:\tin)$-intensity $\Lambda_0^T$ such that $$\E[\int_0^T|\phi(\Lambda_t)|\,dt]<\infty,$$  and let $M$ be a measurable mapping  on the given probability space satisfying $I(M;Y_0^T)<\infty$. Then there exists 
a process $\Gamma_0^T$ such that $\Gamma_0^T$ is the $(\G_t=\sigma(M,Y_0^t):\tin)$ intensity of $Y_0^T$ and 
\begin{align*}
I(M;Y_0^T)=\E\left[\int_0^T \phi(\Gamma_t)-\phi(\Lambda_t)\,dt\right].
\end{align*}
\end{Theorem}

\begin{IEEEproof}
Let $P^{M,Y_0^T}$ denote the joint distribution of $M$ and $Y_0^T$, and $P^M$ and $P^{Y_0^T}$ denote their marginals, respectively. Since $I(M;Y_0^T)<\infty$, we get that $P^{M,Y_0^T}\ll P^M\times P^{Y_0^T}$~\cite[Lemma 5.2.3, p. 92]{Gray}. Lemma~\ref{lem:abs_cont}
says that $P^{Y_0^T} \ll P_0^{Y_0^T}$, which together with~\cite[Chapter 1, Exercise 19, p. 22]{Kallenberg} gives $P^{M,Y_0^T}\ll P^M\times P^{Y_0^T}\ll P^M\times P_0^{Y_0^T}$.

Let $\tilde{P}^{M,Y_0^T}\de{P}^{M}\times P_0^{Y_0^T}$ and
\begin{align}
\mathcal{L}=\frac{d{P}^{M,Y_0^T}}{d\tilde{P}^{M,Y_0^T}}
\label{eq:Lt1}
\end{align}
denote the  Radon-Nikodym derivative. Since under $\tilde{P}^{M,Y_0^T}$, $M$ and $Y_0^T$ are independent, we note that the $(\tilde{P}^{M,Y_0^T},(\G_t:\tin))$-intensity of $Y_0^T$ is 1~\cite[E5 Exercise, Chapter II, p. 28]{Bremaud}. Define the process $L_0^{T}$ as
\begin{align}
L_t=\E_{\tilde{P}}[\mathcal{L}|\G_t], \quad t \in [0, T],
\label{eq:Lt2}
\end{align}
where $\E_{\tilde{P}}$ denotes that the conditional expectation is taken with respect to the measure $\tilde{P}^{M,Y_0^T}$. Then $L_0^T$ is a $(\tilde{P}^{M,Y_0^T},\G_t)$ non-negative absolutely-integrable martingale.

By the martingale representation theorem, the process $L_0^T$ can be written as \cite[Chapter III, Theorem T17, p. 76]{Bremaud} (where we have taken $\sigma(M)$ to be the ``germ $\sigma$-field"):
\begin{align*}
L_t=1+\int_{0}^t K_s(d\Y_s-  ds),
\end{align*}
where $K_0^{T}$ is a $(\G_t:\tin)$-predictable process which satisfies $\int_0^T|K_t|\,dt<\infty$ $\tilde{P}^{M,Y_0^T}$-a.s.
Applying \cite[Lemma 19.5, p. 315]{Liptser}, we can write $L_0^T$ as
\begin{align}
L_t=\exp\left(\int_{0}^t\log (\Gamma_s)\,dY_s+(1-\Gamma_s)\,ds\right),\quad \tin,
\label{EQ:L_T}
\end{align}
where $\Gamma_0^T$ is a non-negative  $(\G_t:\tin)$-predictable process, and $\Gamma_t<\infty$ $\tilde{P}^{M,Y_0^T}$-a.s. for $\tin$.

Now we can mimic the proof of \cite[Chapter VI, Theorems T3, p. 166]{Bremaud} to deduce:
\begin{Lemma}
\label{Le:int_eqn}
For all non-negative $(\G_t:\tin)$-predictable processes
$C_0^T$ 
\begin{align*}
\E\left[\int_0^T C_t\Gamma_t\,dt\right]=\E\left[\int_0^T C_t\,dY_t\right],
\end{align*}
where the expectation is taken with respect to measure $P$.
\end{Lemma}
\begin{IEEEproof}
Please see the supplementary material.
\end{IEEEproof}
 
 Taking $C_t=1$ in the above equality yields  
 \begin{equation}
\E\left [\int_0^T \Gamma_t\,dt\right]=\E\left[\int_0^T \,dY_t\right]=\E\left[\int_0^T \Lambda_t\,dt\right]<\infty.
\label{eq:finite_gammat}
 \end{equation}
 Hence $\int_0^T \Gamma_t\,dt<\infty$ $P$-a.s. and we conclude that the $({P}^{M,Y_0^T},\G_t:\tin)$-intensity of $Y_{0}^{T}$ is $\Gamma_0^{T}$.

Now we will use:
\begin{Lemma}
\label{Le:phi_fun}
\begin{align}
\E\left[\int_{0}^{T}\log (\Gamma_t)\,dY_t\right]=\E\left[\int_{0}^{T}\phi(\Gamma_t)\,dt\right].
\label{eq:phi_fun}
\end{align}
\end{Lemma}
\begin{IEEEproof}
Please see the supplementary material.
\end{IEEEproof}

Since $\E\left[\log\left(\frac{dP^{M,N_{0}^{T}}}{d\tilde{P}^{M,N_{0}^{T}}}\right)\right]$ is  well-defined,  (\ref{eq:Lt1}), (\ref{eq:Lt2}), and (\ref{EQ:L_T})  yields 
\begin{align}
\E\left[\log\left(\frac{dP^{M,N_{0}^{T}}}{d\tilde{P}^{M,N_{0}^{T}}}\right)\right]&=\E\left[\log(L_{T})\right]\nn
&=\E\left[\int_{0}^{T}\log (\Gamma_t)\,dY_t+(1-\Gamma_t)\,dt\right]\nn
&=\E\left[\int_{0}^{T}\phi(\Gamma_t)\,dt\right]+\E\left[ \int_{0}^T(1-\Lambda_t)\,dt\right],
\label{eq:MI-3}
 \end{align}
where in the last line we have used Lemma~\ref{Le:phi_fun} and $\E\left [\int_0^T \Gamma_t\,dt\right]=\E\left[\int_0^T \Lambda_t\,dt\right]<\infty$ from (\ref{eq:finite_gammat}).
Also, 
\begin{align}
\E\left[\log\left(\frac{d(P^{M}\times P^{Y_{0}^{T}}}{d\tilde{P}^{M,Y_{0}^{T}}}\right)\right]&=\E\left[\log\left(\frac{dP^{Y_{0}^{T}}}{dP_0^{Y_{0}^{T}}}\right)\right]\nn
&\overset{(a)}{=}\E\left[\int_{0}^{T}\phi(\Lambda_t) + 1-\Lambda_t\,dt\right] \label{eq:MI-2} \\
&<\infty,\nonumber
\end{align}
where we have used Lemma~\ref{lem:abs_cont} for (a).
Using the above inequality and the fact that $$\E\left[\log\left(\frac{dP^{M,N_{0}^{T}}}{d\tilde{P}^{M,N_{0}^{T}}}\right)\right]$$ is  well-defined, we can express the mutual information as
 \begin{align}
 I(M;Y_0^T)&=\E\left[\log\left(\frac{d{P}^{M,Y_0^T}}{d({P}^{M}\times P^{Y_0^T})}\right)\right]\nn
 &=\E\left[\log\left(\frac{dP^{M,N_{0}^{T}}}{d\tilde{P}^{M,N_{0}^{T}}}\right)\right]-\E\left[\log\left(\frac{d(P^{M}\times P^{N_{0}^{T}}}{d\tilde{P}^{M,N_{0}^{T}}}\right)\right],
 \label{eq:MI-1}
 \end{align}
Now we can compute the mutual information from (\ref{eq:MI-3}), (\ref{eq:MI-2}), and (\ref{eq:MI-1}),  
\begin{align*}
 I(M;Y_0^T)&=\E\left[\int_{0}^{T}\phi(\Gamma_t)\,dt\right]+\E\left[ \int_{0}^T(1-\Lambda_t)\,dt\right]-\E\left[\int_{0}^{T}\phi(\Lambda_t)\,dt\right] - \E\left[\int_{0}^{T}1-\Lambda_t\,dt\right]\nn
 &=\E\left[\int_{0}^{T}\phi(\Gamma_t)\,dt\right]-\E\left[\int_{0}^{T}\phi(\Lambda_t)\,dt\right]\nn
 &=\E\left[\int_{0}^{T}\phi(\Gamma_t)-\phi(\Lambda_t)\,dt\right].
 \end{align*}
\end{IEEEproof}

We shall require several strong data processing inequalities, for which purpose we now derive some ancillary results regarding the intensity of a point process.  
Combining \cite[T8 Theorem, Chapter II, p. 27]{Bremaud} and \cite[T9 Theorem, Chapter II, p. 28]{Bremaud}, we can conclude the following result.
\begin{Lemma}
\label{lem:local_mart}
    Let $\Gamma_0^T$ be a $(\F_t:\tin)$-predictable non-negative process satisfying $$\int_0^T \Gamma_t\,dt<\infty \quad \text{a.s.}$$ Let $Y_0^T$ be a point process adapted to  $(\F_t:\tin)$. Then $\Gamma_0^T$ is the $(\F_t:\tin)$-intensity of  $Y_0^T$  if and only if 
$$
M_t\de Y_t-\int_0^t \Gamma_s\,ds\quad \tin
$$
is a $(\F_t:\tin)$-local martingale\footnote{
    A process $Y_0^T$ is called a \emph{local martingale} with respect to a filtration $(\F_t:t\ge 0)$ if $Y_t$ is $\F_t$-measurable for each $\tin$ and there exists an increasing sequence of stopping times $T_n$, such that $T_n\to\infty$ and the stopped and shifted processes $(Y_{\min\{t,T_n\}}-Y_0:\tin)$ are  $(\F_t:\tin)$-martingales for each $n$.}.
\end{Lemma}
If we impose the stricter condition of finite expectation $\E[\int_0^T \Gamma_t\,dt]<\infty$, the local martingale condition  in the above statement can be replaced by the martingale condition. 

\begin{Lemma}
\label{lem:martingale_int}
Let $\Gamma_0^T$ be a $(\F_t:\tin)$-predictable non-negative process satisfying $$\E\left[\int_0^T \Gamma_t\,dt\right]<\infty.$$ Let $Y_0^T$ be a point process adapted to  $(\F_t:\tin)$. Then $\Gamma_0^T$ is the $(\F_t:\tin)$-intensity of  $Y_0^T$  if and only if 
$$
M_t\de Y_t-\int_0^t \Gamma_s\,ds\quad \tin
$$
is a $(\F_t:\tin)$-martingale.
\end{Lemma}
\begin{IEEEproof}
Please see the supplementary material.
\end{IEEEproof}

\begin{Lemma}
\label{lem:int_his}
If a point process $N_0^T$ has  $(\F_t:\tin)$-intensity ${\Lambda}_0^T$, and $(\G_t:\tin)$ is another history for $N_0^T$ such that 
$\G_t\subseteq \F_t$ for each $\tin$, then there exists a process $\Pi_0^T$ such that $\Pi_0^T$   is the $(\G_t:\tin)$-intensity of ${N}_0^T$, and for each $\tin$, $\Pi_t=\E[{\Lambda}_t|\G_{t-}]$ $P$-a.s.
\end{Lemma}
\begin{IEEEproof}
Please see the supplementary material.
\end{IEEEproof}

\begin{Lemma}
\label{lem:add_int}
Let $Y_0^T$ be a point process with $(\G_t\de\sigma(M,Y_0^t):\tin)$-intensity $\Gamma_0^T$ for some $M$. Let $Z_0^T$ be obtained adding an independent (of both $M$ and $Y_0^T$) point process $N_0^T$ with $(\F_t^N:\tin)$-intensity $\Pi_0^T$ to $Y_0^T$. Then $Z_0^T$ has a $(\F_t\de\sigma(M,Z_0^t):\tin)$-intensity  $ \Theta_0^T$ which satisfies  $\Theta_t=\E[(\Gamma_t+\Pi_t)|\F_{t-}]$ $P$-a.s. for each $\tin$.
\end{Lemma}
\begin{IEEEproof}
Please see the supplementary material.
\end{IEEEproof}

\begin{Theorem}
\label{thm:add_int}
Let $Y_0^T$ be a Poisson process with rate $\lambda$, $M$ be such that $I(M;Y_0^T)<\infty$, and $\Gamma_0^T$ be the $(\sigma(M;Y_0^t):\tin)$-intensity of $Y_0^T$. Suppose $Z_0^T$ is obtained by adding an independent (of $Y_0^T$ and $M$) Poisson process with rate $\mu$ to $Y_0^T$. Then, 
\begin{align*}
I(M;Y_0^T)&=\E\left[\int_0^T \phi(\Gamma_t)-\phi(\lambda)\,dt\right], \nn
I(M;Z_0^T)&\le\E\left[\int_0^T \phi(\Gamma_t+\mu)-\phi(\lambda+\mu)\,dt\right].
\end{align*}
\end{Theorem}
\begin{IEEEproof}
Since $M\leftrightarrows Y_0^T \leftrightarrows Z_0^T$ forms a Markov chain, the data processing inequality gives $I(M;Z_0^T)\le I(M;Y_0^T)<\infty$. Applying Theorem~\ref{thm:MI} and using the uniqueness of intensities,
\begin{align}
I(M;Y_0^T)&=\E\left[\int_0^T \phi(\Gamma_t)-\phi(\lambda)\,dt\right],\quad\text{and}\nn
I(M;Z_0^T)&=\E\left[\int_0^T \phi(\hat{\Gamma}_t)-\phi(\hat{\lambda}_t)\,dt\right],
\label{eq:MI_Z2}
\end{align}
where $\hat{\Gamma}_0^T$  and $\hat{\lambda}_0^T$ are the $(\sigma(M;Z_0^t):\tin)$ and $(\F_t^Z:\tin)$-intensities  of $Z_0^T$.
Due to the uniqueness of the intensities and  Lemma~\ref{lem:add_int}, we get for each $\tin$, $\hat{\Gamma}_t=\E[\Gamma_t|M,Z_0^{t-}]+\mu$, and $\hat{\lambda}_t=\lambda+\mu$. Substituting this in (\ref{eq:MI_Z2}) and applying Jensen's inequality yields
\begin{align*}
I(M;Z_0^T)&=\E\left[\int_0^T \phi(\E[\Gamma_t|M,Z_0^{t-}]+\mu)-\phi(\lambda+\mu)\,dt\right],\\
&\le\E\left[\int_0^T \phi(\Gamma_t+\mu)-\phi(\lambda+\mu)\,dt\right].
\end{align*}
\end{IEEEproof}

\begin{Def}
A point process $Z_0^T$ is said to be obtained from $p$-thinning of a point process $Y_0^T$, if each point in $Y_0^T$ is deleted with probability $p$, independent of all other points and deletions. 
\end{Def}
\begin{Lemma}
\label{lem:thin_int}
Suppose that $Y_0^T$ is a point process with $\G_t\de\sigma(M,Y_0^t)$-intensity $\Gamma_0^T$ such that $\E[\int_0^T \Gamma_t\,dt]<\infty$ and $Z_0^T$ is obtained from $p$-thinning $Y_0^T$.  Then the $(\F_t\de\sigma(M,Z_0^t):\tin)$-intensity of $Z_0^T$ is given by $\Theta_0^T$, where $P$-a.s. $\Theta_t =(1-p)\E[\Gamma_t|\F_{t-}], \tin$.
\end{Lemma}
\begin{IEEEproof}
Please see the supplementary material.
\end{IEEEproof}

The following theorem was first proven  by Wang in~\cite{Wang'17} using a property of certain ``contraction coefficient" used in strong data processing inequalities~\cite{Polyanskiy'17}. Here, we provide a self-contained proof which uses Theorem~\ref{thm:MI} and Lemma~\ref{lem:thin_int}.  
\begin{Theorem}
\label{thm:thin_int}
Let $Y_0^T$ be a Poisson process with rate $\lambda$, and $M$ be such that $I(M;Y_0^T)<\infty$. Let $Z_0^T$ obtained from $p$-thinning of $Y_0^T$ such that   the thinning operation is independent of $M$. Then
\begin{align*}
I(M;Z_0^T) \le (1-p)I(M;Y_0^T).
\end{align*}
\end{Theorem}

\begin{IEEEproof}
The data processing inequality gives $I(M;Z_0^T)\le I(M;Y_0^T)<\infty$. Applying Theorem~\ref{thm:MI},
\begin{align}
I(M;Y_0^T)=\E\left[\int_0^T \phi(\Gamma_t)-\phi(\lambda)\,dt\right],
\end{align}
and 
\begin{align}
I(M;Z_0^T)=\E\left[\int_0^T \phi(\hat{\Gamma}_t)-\phi(\hat{\lambda}_t)\,dt\right],
\label{eq:MI_Z}
\end{align}
where $\Gamma_0^T$ and $\lambda_0^T$ (respectively $\hat{\Gamma}_0^T$ and $\hat{\lambda}_0^T$) are the $(\sigma(M;Y_0^t):\tin)$ and $(\sigma(Y_0^t):\tin)$-intensities (respectively $(\sigma(M;Z_0^t):\tin)$ and $(\sigma(Z_0^t):\tin)$-intensities) of $Y_0^T$ (respectively $Z_0^T$).
Due to the uniqueness of the intensities and  Lemma~\ref{lem:thin_int},  we can take for each $\tin$, 
$$\hat{\Gamma}_t=(1-p)\E[\Gamma_t|M,Z_0^{t-}],\quad\hat{\lambda}_t=(1-p)\lambda.$$
Noting that $\phi((1-p)x)=(1-p)\phi(x)+x\phi(1-p)$, (\ref{eq:MI_Z}) yields  
\begin{align*}
I(M;Z_0^T)=&(1-p)\E\left[\int_0^T \phi(\E[\Gamma_t|M,Z_0^{t-}])-\phi(\lambda)\,dt\right] \nn
&+\phi(1-p)\E\left[\int_0^T\Gamma_t-\lambda \,dt \right] \nn
\overset{(a)}{=}&(1-p)\E\left[\int_0^T \phi(\E[\Gamma_t|M,Z_0^{t-}])-\phi(\lambda)\,dt\right]\\
\overset{(b)}{\le}&(1-p)\E\left[\int_0^T \phi(\Gamma_t)-\phi(\lambda)\,dt\right]\nn
=&(1-p)I(M;Y_0^T),
\end{align*}
where for (a) we have used the fact that $\E\left[ \int_{0}^T \Gamma_t\,dt\right]=\E\left[ \int_{0}^T 1\,dY_t\right]=\E\left[ \int_{0}^T \lambda\,dt\right]$, and \\*
for (b) we have used Jensen's inequality. 
\end{IEEEproof}

We will  require the following result~\cite[Theorem 2.11, p. 106]{durrett2016essentials}.
\begin{Lemma}
\label{le:Poiss_Split}
Suppose that $Y_0^T$ is a Poisson process with rate $\lambda$ and $Z_0^T$ is obtained from $p$-thinning of $Y_0^T$. Let
\begin{align*}
\hat{Z}_t=Y_t-Z_t \quad \tin.
\end{align*}
Then $\hat{Z}_0^T$ and $Z_0^T$ are independent Poisson processes with rates $p\lambda$ and $(1-p)\lambda$ respectively.
\end{Lemma}

The following lemma will be used repeatedly in the converse proofs of the rate-distortion function.
\begin{Lemma}
\label{Le:hat_Yt}
Let a point process $Y_0^T$ have an $(\F_t:\tin)$-intensity $\Gamma_0^T$ such that $$\E\left[\int_0^T\phi(\Gamma_t)\,dt\right]<\infty.$$ Let $\hat{Y}_0^T$ be an non-negative $(\F_t:\tin)$-predictable process satisfying $\E\left[\int_0^T \hat{Y}_t\,dt\right]<\infty$. Then
\begin{align*}
\E\left[\int_0^T \log(\hat{Y}_t)\,dY_t \right]=\E\left[\int_0^T \log(\hat{Y}_t)\Gamma_t\,dt \right].
\end{align*}
\end{Lemma}
\begin{IEEEproof}
Please see the supplementary material.
\end{IEEEproof}

\section{Functional Covering of Point Processes}
In this section, we will consider general point processes  and obtain the rate-distortion function under the functional-covering distortion when feedforward is present. Stronger results are obtained for Poisson processes in the next sections.  
\begin{Def}
Given a point process $y_0^T\in\mathcal{N}_0^T$, and a non-negative function $\hat{y}_0^T$, the \emph{functional-covering distortion} $d$ is   
\begin{align*}
d(\hat{y}_0^T,y_0^T)\de\left(\int_0^T \hat{y}_t\,dt-\log(\hat{y}_t)\,dy_t \right),
\end{align*}
whenever the expression on the right is well-defined.
\end{Def}

We will allow the reconstruction function $\hat{Y}_0^T$ to depend on $Y_0^T$ as well as the message, constrained via predictability. In particular, we will call $\hat{Y}_0^T$ an \emph{allowable reconstruction with feedforward} if it is non-negative and $(\sigma(Y_0^t):\tin)$-predictable. Let $\mathcal{\hat{Y}}_{0,\text{FF}}^T$ denote the set of all  $\hat{y}_0^T$ processes which are allowable reconstructions with feedforward.
\begin{Def}
\label{def:dist1}
A \emph{$(T,R,D)$ code with feedforward} consists of an \emph{encoder} $f$
\begin{align*}
f:\mathcal{N}_0^T \rightarrow \{1,\dots,\dots,\lceil\exp(RT)\rceil\}
\end{align*}
and a \emph{decoder} $g$
\begin{align*}
g:\{1,\dots,\lceil\exp(RT)\rceil\}\times \mathcal{N}_0^T \rightarrow \mathcal{\hat{Y}}_{0,\text{FF}}^T 
\end{align*}
satisfying
\begin{align*}
\E\left[\int_0^T \hat{Y}_t\,dt\right]<\infty
\end{align*}
and the distortion constraint 
\begin{align*}
\E\left[\frac{1}{T}d(\hat{Y}_0^T,Y_0^T)  \right] \le D.
\end{align*}
\end{Def}
We will call the encoder's output $M=f(Y_0^T)$  the \emph{message} and the decoder's output $\hat{Y}_0^T$  the \emph{reconstruction.}
\begin{Def}
The \emph{minimum achievable distortion with feedforward at rate $R$ and blocklength $T$} is 
\begin{align*}
D_F^*(R,T)\de\inf\{D: \text{there exists a $(T,R,D)$ code with feedforward} \}.
\end{align*}
\end{Def}
\begin{Def}
The  \emph{distortion-rate function with feedforward} is
\begin{align*}
D_F(R)\de\limsup_{T\to\infty}D_F^*(R,T).
\end{align*}
\end{Def}
The minimum achievable rate at distortion $D$ and blocklength $T$ with feedforward $R_F^*(D,T)$ and the rate-distortion function with feedforward $R_F(D)$ can be defined similarly.

$D_F^*(R,T)$ can be characterized via the following theorem for certain point processes. 

\begin{Theorem}
\label{thm:R-D}
Let $Y_0^T$ be a point process with $(\F_t^Y:\tin)$-intensity $\Lambda_0^T$ such that $$\E\left[\int_0^T|\phi(\Lambda_t)|\,dt\right]<\infty.$$ Let 
\begin{align*}
\Xi(Y_0^T)\de\frac{1}{T}\E\left[\int_0^T \Lambda_t-\phi(\Lambda_t)\,dt \right],
\end{align*}
and $$\delta_T\de P(Y_T=0)<1.$$ 
Then $D_F^*(R,T)$ satisfies
\begin{align*}
\Xi(Y_0^T)-R-\frac{1}{T} \le D_F^*(R,T) \le  \Xi(Y_0^T)-(1-\delta_T)R+\frac{1}{T}.
\end{align*}
\end{Theorem}
\begin{IEEEproof}\\
\emph{Achievability:}\\
Recall that since $\Lambda_0^T$ is the $(\F_t^Y:\tin)$-intensity of  $Y_0^T$, it is $(\F_t^Y:\tin)$-predictable, and  $\E[\int_0^T|\phi(\Lambda_t)|\,dt]<\infty$ implies $\E[\int_0^T\Lambda_t\,dt]<\infty$.
If the decoder outputs $\Lambda_0^T$, this leads  to distortion 
\begin{align*}
\frac{1}{T}\E[d(\Lambda_0^T,Y_0^T)]&=\frac{1}{T}\E\left[\int_0^T \Lambda_t\,dt-\log(\Lambda_t)\,dY_t\right]\\
&=\frac{1}{T}\E\left[\int_0^T \Lambda_t-\phi(\Lambda_t)\,dt \right]\\
&= \Xi(Y_0^T).
\end{align*}
Thus $D_F^*(0,T)\le\Xi(Y_0^T)$, and the upper bound in the statement of the theorem holds at $R=0$.

Now consider the case $R>0$. Fix $T>0$ and let $J=\lceil\exp(RT)\rceil$.
 If $Y_T=0$, then the encoder sends index $M=1$. Otherwise, let $\Theta$ denote the first arrival instant of the observed point process  $Y_0^T$. 
From Lemma~\ref{lem:abs_cont}, we have that $P^{Y^T_0}\ll P_0^{Y^T_0}$. Since under $P_0^{Y^T_0}$, $Y_0^T$ is a Poisson process with unit rate, it holds that  $P_0^{Y^T_0}(\Theta=t, Y_T>0)=0$ for any fixed $\tin$. This gives us $P(\Theta=t, Y_T>0)=0$ for $\tin$. Thus conditioned on the event $Y_T> 0$,  $\Theta$ has a continuous distribution function $F_{\Theta}$.
The encoder computes $F_{\Theta}(\Theta)$ which is uniformly distributed over $[0,1]$, which the encoder suitably quantizes to obtain an $M$ which is uniform in $\{2,\dots,J\}$.  From Theorem~\ref{thm:MI}, there exists a $(\sigma(M,Y_0^t):\tin)$-predictable process $\Gamma_0^T$ which is the $(\sigma(M,Y_0^t):\tin)$-intensity of $Y_0^T$. We note that $\E\left[\int_0^T \Gamma_t\,dt\right]=\E\left[\int_0^T \Lambda_t\,dt\right]<\infty$, and 
from Theorem~\ref{thm:MI}, $\E\left[\int_0^T\log(\Gamma_t)\,dY_t\right]<\infty$. Hence
$$
\frac{1}{T}\E[d(\Lambda_0^T,Y_0^T)]=\frac{1}{T}\E\left[\int_0^T \Gamma_t\,dt-\log(\Gamma_t)\,dY_t\right]
$$
is well-defined. The decoder outputs $\Gamma_0^T$ as its reconstruction.
Then we have
\begin{align}
\frac{1}{T}H(M)&=-\frac{1}{T}\left(\delta_T\log(\delta_T)+(1-\delta_T)\log(1-\delta_T)\right)+\frac{1-\delta_T}{T}(\log(J-1))\nn
&\overset{(a)}{\ge}  \frac{1-\delta_T}{T}(\log(J-1))\nn
&\overset{(b)}{\ge}   \frac{1-\delta_T}{T}\log(J/\exp(1))\nn
&\overset{(c)}{\ge}  (1-\delta_T)R -\frac{1}{T},
\label{eq:HM_bound2}
\end{align}
where for (a), we have used the bound $-\delta_T\log(\delta_T)-(1-\delta_T)\log(1-\delta_T)\ge 0$,\\*
for (b), we have used the inequality $J-1\ge J/\exp(1)$ when $J\ge 2$, and \\*
for (c), we used the fact that  $ RT\le \log(J)$.

$H(M)$ also satisfies
\begin{align}
\frac{1}{T}H(M)&\overset{(a)}{=}\frac{1}{T}I(M;Y_0^T)\nn
&\overset{(b)}{=}\frac{1}{T}\E\left[\int_0^T \log(\Gamma_t)\,dY_t\right]-\frac{1}{T}\E\left[\int_0^T \phi(\Lambda_t)\,dt\right],
\label{eq:HM_bound1}
\end{align}
where, for (a) we have used Lemma~\ref{Le:Wyner},\\*
for (b) we have used Theorem~\ref{thm:MI}.\\*
The average distortion can be bounded as follows:
\begin{align*}
\frac{1}{T}\E[d(\Lambda_0^T,Y_0^T)]&=\frac{1}{T}\E\left[\int_0^T \Gamma_t\,dt-\log(\Gamma_t)\,dY_t\right]\\
&\overset{(a)}{=}\frac{1}{T}\E\left[\int_0^T \Gamma_t\,dt\right]-\frac{1}{T}\E\left[\int_0^T\log(\Gamma_t)\,dY_t\right]\\
&\overset{(b)}{=}\frac{1}{T}\E\left[\int_0^T \Lambda_t\,dt\right]-\frac{1}{T}\E\left[\int_0^T\log(\Gamma_t)\,dY_t\right]\\
&\overset{(c)}{=}\frac{1}{T}\E\left[\int_0^T \Lambda_t\,dt\right]-\frac{1}{T}H(M)-\frac{1}{T}\E\left[\int_0^T \phi(\Lambda_t)\,dt\right]\\
&\overset{(d)}{\le} \frac{1}{T}\E\left[\int_0^T \Lambda_t-\phi(\Lambda_t)\,dt\right] -(1-\delta_T)R+\frac{1}{T}\\
&= \Xi(Y_0^T)-(1-\delta_T)R+\frac{1}{T},
\end{align*}
where, for (a), we have used the fact that  $\E\left[\int_0^T\log(\Gamma_t)\,dY_t\right]<\infty$ due to Theorem~\ref{thm:MI},\\*
for  (b), we used the equality $\E\left[\int_0^T \Gamma_t\,dt\right]=\E\left[\int_0^T \Lambda_t\,dt\right]$,\\*
for (c), we used~(\ref{eq:HM_bound1}), and \\*
for (d), we used~(\ref{eq:HM_bound2}).

Thus we have shown the existence of a $(T,R,D)$ code with feedforward such that $D=\Xi(Y_0^T)-(1-\delta_T)R+\frac{1}{T}$. This gives the upper bound on $D_F^*(R,T)$.
\subsection*{Converse:}
For the given $(T,R,D)$ code with feedforward, let $J=\lceil\exp(RT)\rceil$. Then $J\le\exp(RT)+1\le \exp(RT+1)$.  Thus we have
\begin{align}
R+\frac{1}{T}\ge\frac{1}{T} \log(J) \ge \frac{1}{T}H(M) \overset{(a)}{=} \frac{1}{T}I(M;Y_0^T),
\label{eq:rd-conv_r}
\end{align}
where (a) follows because of Lemma~\ref{Le:Wyner}.

Since $I(M;Y_0^T)<\infty$, we conclude from Theorem~\ref{thm:MI} that there exists 
a process $\Gamma_0^T$ such that $\Gamma_0^T$ is the $(\F_t=\sigma(M,Y_0^t):\tin)$ intensity of $Y_0^T$ and 
\begin{align*}
I(M;Y_0^T)=\E\left[\int_0^T \phi(\Gamma_t)\,dt\right]-\E\left[\int_0^T\phi(\Lambda_t)\,dt\right].
\end{align*}
Hence from~(\ref{eq:rd-conv_r})
\begin{align}
R \ge \frac{1}{T}\E\left[\int_0^T \phi(\Gamma_t)\,dt\right]-\frac{1}{T}\E\left[\int_0^T\phi(\Lambda_t)\,dt\right]-\frac{1}{T}.
\label{eq:RD_conv1}
\end{align}

Let $\hat{Y}_0^T$ denote the decoder's output. The distortion constraint $D$ satisfies
\begin{align}
D\ge \frac{1}{T}\E\left[d(\hat{Y}_0^T,Y_0^T) \right]&=\frac{1}{T}\E\left[\int_0^T \hat{Y}_t\,dt-\log(\hat{Y}_t)\,dY_t\right]\nn
&=\frac{1}{T}\E\left[\int_0^T \hat{Y}_t-\log(\hat{Y}_t)\Gamma_t\,dt\right]
\label{eq:rd-conv-d}
\end{align}
where in the last line we have used Lemma~\ref{Le:hat_Yt}.

Using the inequality $u\log(v)\le \phi(u)-u+v$, and noting that the individual terms have finite expectations,
\begin{align}
\E\left[\int_0^T \log(\hat{Y}_t)\Gamma_t\,dt \right]&\le \E\left[\int_0^T \phi(\Gamma_t) - \Gamma_t+ \hat{Y}_t\,dt\right]\nn
&=\E\left[\int_0^T \phi(\Gamma_t)\,dt\right] - \E\left[\int_0^T\Gamma_t\,dt\right]+ \E\left[\int_0^T\hat{Y}_t\,dt\right]
 \label{eq:RD_conv3_gen}.
\end{align}

From (\ref{eq:rd-conv-d}) and (\ref{eq:RD_conv1}), we deduce
\begin{align*}
R+D&\ge\frac{1}{T}\E\left[\int_0^T \phi(\Gamma_t)\,dt\right]-\frac{1}{T}\E\left[\int_0^T\phi(\Lambda_t)\,dt\right]+\frac{1}{T}\E\left[\int_0^T \hat{Y}_t\,dt\right]\\
      & \phantom{====} -\frac{1}{T}\E\left[\int_0^T\log(\hat{Y}_t)\,dY_t\right]-\frac{1}{T}\\
&\overset{(a)}{\ge}\frac{1}{T}\E\left[\int_0^T\Gamma_t\,dt\right]-\frac{1}{T}\E\left[\int_0^T \phi(\Lambda_t)\,dt\right]-\frac{1}{T}\\
&\overset{(b)}{\ge}\frac{1}{T}\E\left[\int_0^T\Lambda_t\,dt\right]-\frac{1}{T}\E\left[\int_0^T \phi(\Lambda_t)\,dt\right]-\frac{1}{T}\\
&\overset{}{=} \Xi(Y_0^T)-\frac{1}{T},
\end{align*}
where, for (a) we have used~(\ref{eq:RD_conv3_gen}), and \\* 
for (b) we used the fact that $\E\left[\int_0^T\Gamma_t\,dt\right]=\E\left[\int_0^T\,dY_t\right]=\E\left[\int_0^T\Lambda_t\,dt\right]$.\\*
Hence we have shown that for any $(T,R,D)$ code with feedforward, $D\ge\Xi(Y_0^T)-R-1/T$. This gives us the lower bound on $D_F^*(R,T)$
\end{IEEEproof}

\begin{Corollary}
\label{cor:R-D}
Let $Y_0^T$ be a point process with $(\F_t^Y:\tin)$-intensity $\Lambda_0^T$ such that 
\begin{itemize}
\item $\E[\int_0^T|\phi(\Lambda_t)|\,dt]<\infty$,
\item $\bar{\Xi}(Y)\de\limsup_{T\to\infty}\frac{1}{T}\E\left[\int_0^T \Lambda_t-\phi(\Lambda_t)\,dt \right]$ is finite.
\item $\lim_{T\to\infty}P(Y_T=0)=0$.
\end{itemize}
Then
\begin{align*}
D_F(R)=\bar{\Xi}(Y)-R.
\end{align*}
\end{Corollary}
\begin{IEEEproof}
The corollary follows from the definition $D_F(R)=\limsup_{T\to\infty}D_F^*(R,T)$ and from the bounds on $D_F^*(R,T)$ in the Theorem~\ref{thm:R-D}.
\end{IEEEproof}
\begin{Remark}
\label{rem:thmR_D}
The above distortion-rate function is reminiscent of the logarithmic-loss distortion-rate function for a DMS. Specifically, for a DMS $Y$ on alphabet $\mathcal{Y}$  let the reconstruction be a probability distribution function $Q$ on  $\mathcal{Y}$. The logarithmic loss distortion is  defined as $d_{LL}(y,Q)\de -\log(Q(y))$ and  the distortion-rate function is then given by $D(R)=(H(Y)-R)^+$~\cite{Courtade2014}. 

If the reconstruction $\hat{y}_0^T$ is assumed to be bounded then it can be used define a probability measure on the space of point-processes $(\mathcal{N}_0^T,\Fk^N)$ via following Radon-Nikodym derivative.
\[
\frac{dP_{\hat{y}_0^T}}{dP_0}(y_0^T) = \exp\left(\int_0^T\log( \hat{y}_t)\,dy_t-(\hat{y}_t-1)\,dt \right),
\]
where $P_0$ is the measure under which $Y_0^T$ is a Poisson process with unit rate.
Then the intensity of $Y_0^T$ under this measure is $\hat{y}_0^T$~\cite[Chapter VI, Theorems T2-T4]{Bremaud} and the functional-covering distortion is related to the above Radon-Nikodym derivative as 
\[
d(\hat{y}_0^T,y_0^T) = -\log\left(\frac{dP_{\hat{y}_0^T}}{dP_{0}} (y_0^T)\right)+T.
\]
\eor
\end{Remark}

Applying the above corollary to a Poisson process with rate $\lambda>0$, we get that $D_F(R)=\lambda-\lambda\log(\lambda)-R$. As we will see in the next section, this distortion-rate function can be achieved  without feedforward.

\section{Constrained Functional-Covering of Poisson Processes}
In this and the next section we focus on Poisson processes. Let $\hat{\mathcal{Y}}_0^T$ denote the set of all functions $\hat{y}_0^T$ which are non-negative and left-continuous with right-limits. We assume that we are given a set $\A\in\mathbb{R}_{+}$ with at least one positive element. We will constrain the reconstruction function $\hat{Y}_0^T$ to take value in $\mathcal{A}$,  so that for all $\tin$, $\hat{Y}_t\in\A$. 

\begin{Def}
A $(T,R,D)$ code consists of an \emph{encoder} $f$
\begin{align*}
f:\mathcal{N}_0^T \rightarrow \{1,\dots,\lceil\exp(RT)\rceil \}
\end{align*}
and a \emph{decoder} $g$
\begin{align*}
g:\{1,\dots,\lceil\exp(RT)\rceil\} \rightarrow \hat{\mathcal{Y}}_0^T
\end{align*}
satisfying
\begin{align*}
\hat{Y}_t\in\A, \, \E\left[\int_0^T \hat{Y}_t\,dt\right]<\infty
\end{align*}
and the distortion constraint 
\begin{align*}
\frac{1}{T}\E\left[d(\hat{Y}_0^T,Y_0^T)  \right] \le D.
\end{align*}
\end{Def}
As before, we will call the encoder's output $M=f(Y_0^T)$  the \emph{message} and the decoder's output $\hat{Y}_0^T=g(M)$  the \emph{reconstruction.}
\begin{Def}
\label{def:RD_cons1}
A {rate-distortion vector} $(R,D)$ is said to be \emph{achievable}  if for any $\epsilon>0$, there exists a sequence of $(T_n,R+\epsilon,D+\epsilon)$ codes such that $\lim_{n\to\infty}T_n=\infty$.
\end{Def}
\begin{Def}
\label{def:RD_cons2}
The \emph{rate-distortion region} $\mathfrak{RD}^{\mathcal{P}}_{\A}$ is the intersection of all achievable rate-distortion vectors $(R,D)$.
\end{Def}


The rate-distortion region $\mathfrak{RD}^{\mathcal{P},\text{F}}_{\A}$ with feedforward is defined as in Definitions~\ref{def:RD_cons1} and \ref{def:RD_cons2}.
\begin{Theorem}
\label{thm:Poisson_Cons}
The rate-distortion region for the constrained functional-covering of a Poisson process with rate $\lambda>0$ is given by
$$\mathfrak{RD}^{\mathcal{P}}_{\A}=\mathfrak{RD}^{\mathcal{P},\text{F}}_{\A}=\mathfrak{RD},$$ where $\mathfrak{RD}$ is the  convex hull of the union of sets of rate-distortion vectors  $(R,D)$ such that 
\begin{gather*}
R \ge \lambda \sum_{k=1}^4 \beta_{k}\log\left(\frac{\beta_{k}}{\alpha_{k}}\right) \\
D \ge \sum_{k=1}^4  \alpha_k \Psi_\A\left(\frac{\lambda\beta_k}{\alpha_k}\right),
\end{gather*}
where 
\begin{align*}
\Psi_\A(u)\de \inf_{v\in\A}v-u\log(v)
\end{align*}
with the convention that $0\Psi(0/0)=0$,
and  $[\alpha_{k}]_{k=1}^4$ and $[\beta_{k}]_{k=1}^4$ are probability vectors over $\{1,2,3,4\}$ satisfying  $\alpha_{k}=0 \Rightarrow \beta_{k}=0$.
\end{Theorem}
\begin{IEEEproof}
\paragraph*{Achievability}
Let 
\begin{gather*}
{R}\de \lambda \sum_{k=1}^4 \beta_{k}\log\left(\frac{\beta_{k}}{\alpha_{k}}\right) \\
{D} \de \sum_{k=1}^4  \alpha_k \Psi_\A\left(\frac{\lambda\beta_k}{\alpha_k}\right).
\end{gather*}
We will show achievability using a $(T,{R}+\epsilon,{D}+\epsilon)$ code without feedforward.  We will use discretization and results from the rate-distortion theory for discrete memoryless sources (DMS). 
Define a binary-valued discrete-time process $(\bar{Y}_{j}:j\in\{1,\dots,n\})$ as follows. If there are one or more arrivals in the interval $((j-1)\Delta, j\Delta]$ of the process $Y_{0}^T$ , then set $\bar{Y}_{j}$ to $1$, otherwise it equals zero. Since $Y_{0}^T$ is a Poisson process with rate $\lambda$, the components of $(\bar{Y}_{j}:j\in\{1,\dots,n\})$ are independent and identically distributed with $P(\bar{Y}=1)=1-\exp(-\lambda\Delta)$.
Consider  the following ``test"-channel for $k\in\{1,2,3,4\}$,
    \begin{align*}
        P(\bar{U}=k|\bar{Y}=1) & =\beta_{k}, \\
        P(\bar{U}=k|\bar{Y}=0) & = \alpha_{k}.
    \end{align*}
Define the discretized distortion function
    \begin{equation*}
\bar{d}(\hat{\bar{y}},\bar{y})\de \hat{\bar{y}}-\frac{\log(\hat{\bar{y}})}{\Delta}\1\{\bar{y}=1\} \quad \hat{\bar{y}}\in\A, \bar{y}\in\{0,1\}.
    \end{equation*}
The reconstruction $ \hat{\bar{Y}}(k)$ is taken as a $v\in\A$ satisfying
\begin{align}
\left|\Psi_\A\left(\frac{\lambda\beta_k}{\alpha_k}\right)-\left(v-\frac{\lambda\beta_k}{\alpha_k}\log(v) \right)\right|\le\frac{\epsilon}{4},
\label{eq:CFC_Psi}
\end{align}
where such a $v$ exists due to the definition of $\Psi_\A$. We recall that if $\alpha_k=0$ then $\beta_k=0$, and hence $P(\bar{U}=k)=0$ for such a $k$.
The scaling of the mutual information $I(\bar{U};\bar{Y})$ and the distortion function $\bar{d}(\hat{\bar{Y}},\bar{Y})$ with respect to $\Delta$ is given by the following lemma.
\begin{Lemma}
\begin{align*}
\lim_{\Delta\to 0}\frac{I(\bar{U};\bar{Y})}{\Delta}&={R}\\
\lim_{\Delta\to 0}\E[\bar{d}(\hat{\bar{Y}},\bar{Y})]&\le{D}+\frac{\epsilon}{4}
\end{align*}
\label{le:delta_scaling_cons}
\end{Lemma}
\begin{IEEEproof}
Please see the supplementary material.
\end{IEEEproof}

Let
\begin{align}
\kappa\de\max_{\substack{\kin\\\hat{\bar{Y}}(k)>0}}\left|\log\left(\hat{\bar{Y}}(k)\right)\right|.
\label{eq:kappa_cons}
\end{align}
Due to~\cite[Theorem 9.3.2, p. 455]{Gallager}, for a given $\Delta>0$, $\bar{\epsilon}>0$, and all sufficiently large $n$, there exists an encoder $\bar{f}$ and a decoder $\bar{g}$ such that 
\begin{gather*}
\bar{f}:(\bar{Y}_{j}:j\in\{1,\dots,n\})\to \{1,\dots,L\}\\
\bar{g}:\{1,\dots,L\}\to(\hat{\bar{Y}}_{j}:j\in\{1,\dots,n\})
\end{gather*}
satisfying
\begin{align}
\frac{1}{n}\log(L)\le I(\bar{U};\bar{Y})+\bar{\epsilon},\nn
\E\left[\frac{1}{n}\sum_{j=1}^n\bar{d}(\hat{\bar{Y}}_j,\bar{Y_j})\right] \le \E[\bar{d}(\hat{\bar{Y}},\bar{Y})]+\bar{\epsilon}.
\label{eq:dis_code_c}
\end{align}

Given the above setup, the encoder $f$ upon observing $Y_0^{T}$ obtains  the binary valued discrete time process $(\bar{Y}_{j}:j\in\{1,\dots,n\})$, and sends $M=\bar{f}(\bar{Y}_{j}:j\in\{1,\dots,n\})$ to the decoder. The decoder outputs the reconstruction $\hat{Y}_0^T$ as
\begin{align*}
\hat{Y}_t\de \sum_{j=1}^n\hat{\bar{Y}}_j\1\left\{t\in((j-1)\Delta, j\Delta]\right\} \quad \tin.
\end{align*}
Let $\bar{\bar{Y}}_j$ denote the actual number of arrivals of $Y_0^T$ in an interval $((j-1)\Delta,j\Delta]$. Then $\bar{d}$ is related to  the original distortion function via the above reconstruction as follows:
\begin{align*}
\frac{1}{T}d(\hat{Y}_0^T;Y_0^T)&=\frac{1}{T}\int_{0}^T\hat{Y}_t\,dt-\frac{1}{T}\int_{0}^T\log(\hat{Y}_t)\,dY_t\\
&=\frac{1}{n}\sum_{j=1}^{n}\hat{\bar{Y}}_j-\frac{1}{T}\sum_{j=1}^{n}\log(\hat{\bar{Y}}_j)\bar{\bar{Y}}_j\\
&=\frac{1}{n}\sum_{j=1}^{n}\hat{\bar{Y}}_j-\frac{1}{n\Delta}\sum_{j=1}^{n}\log(\hat{\bar{Y}}_j)\bar{Y}_j-\frac{1}{T}\sum_{j=1}^{n}\log(\hat{\bar{Y}}_j)(\bar{\bar{Y}}_j-1)\1\{\bar{\bar{Y}}_j>1\}.
\\&= \frac{1}{n}\sum_{j=1}^{n} \bar{d}(\hat{\bar{Y}}_j,\bar{Y}_j)-\frac{1}{T}\sum_{j=1}^{n}\log(\hat{\bar{Y}}_j)(\bar{\bar{Y}}_j-1)\1\{\bar{\bar{Y}}_j>1\}\\
&\overset{(a)}{\le} \frac{1}{n}\sum_{j=1}^{n} \bar{d}(\hat{\bar{Y}}_j,\bar{Y}_j)+\frac{\kappa}{T}\sum_{j=1}^{n}(\bar{\bar{Y}}_j-1)\1\{\bar{\bar{Y}}_j>1\}\\
&\le \frac{1}{n}\sum_{j=1}^{n} \bar{d}(\hat{\bar{Y}}_j,\bar{Y}_j)+\frac{\kappa}{T}\sum_{j=1}^{n}\bar{\bar{Y}}_j\1\{\bar{\bar{Y}}_j>1\},
\end{align*}
where for (a), we have used the definition of $\kappa$ in~(\ref{eq:kappa_cons}), since $\bar{\bar{Y}}_j>1$ implies $\bar{Y}_j=1$ which implies $\hat{\bar{Y}}_j>0$ in order for $\bar{d}(\hat{\bar{Y}}_j,1)<\infty$, which occurs a.s. since $\E[\bar{d}(\hat{\bar{Y}},\bar{Y})]<\infty$ so long as $\Delta$ is sufficiently small. \\*
Hence taking expectations, we get 
\begin{align}
\E\left[\frac{1}{T}d(\hat{Y}_0^T,Y_0^T)\right]&
\le \E\left[\frac{1}{n}\sum_{j=1}^{n} \bar{d}(\hat{\bar{Y}}_j,\bar{Y}_j)\right]+\kappa\E\left[\frac{1}{T}\sum_{j=1}^{n}\bar{\bar{Y}}_j\1\{\bar{\bar{Y}}_j>1\}\right]\nn
&\overset{(a)}{\le}\E[\bar{d}(\hat{\bar{Y}},\bar{Y})]+\kappa\E\left[\frac{1}{T}\sum_{j=1}^{n}\bar{\bar{Y}}_j\1\{\bar{\bar{Y}}_j>1\}\right]+\bar{\epsilon}\nn
&\overset{(b)}{=}\E[\bar{d}(\hat{\bar{Y}},\bar{Y})]+\kappa(\lambda-\lambda\exp(-\lambda\Delta))+\bar{\epsilon}\nn
&\overset{(c)}{\le} \E[\bar{d}(\hat{\bar{Y}},\bar{Y})]+\kappa\lambda^2\Delta+\bar{\epsilon},
\label{eq:cons_ach_dbound1}
\end{align}
where, for (a), we have used~(\ref{eq:dis_code_c}),\\*
for (b) we note that $\E[\bar{\bar{Y}}_j\1\{\bar{\bar{Y}}_j>1\}]=\lambda\Delta-\lambda\Delta\exp(-\lambda\Delta)$, and \\*
for (c), we have used the inequality $1-u\le \exp(-u)$ .\\*
Moreover using~(\ref{eq:dis_code_c}), 
\begin{align}
\frac{1}{T}\log(L)=\frac{1}{n\Delta}\log(L)\le \frac{I(\bar{U};\bar{Y})}{\Delta}+\frac{\bar{\epsilon}}{\Delta}.
\label{eq:code-size_c}
\end{align}

Now given the rate-distortion vector $({R},{D})$ and $\epsilon>0$, first choose $\Delta<1$ sufficiently small so that 
\begin{align*}
\frac{I(\bar{U};\bar{Y})}{\Delta}&\le{R}+\frac{\epsilon}{4}\\
\E[\bar{d}(\hat{\bar{Y}},\bar{Y})]&\le{D}+\frac{\epsilon}{2},\\
\kappa\lambda^2\Delta&\le\epsilon/2.
\end{align*}
Then let $\bar{\epsilon}=\Delta\epsilon/4$, and choose a sufficiently large $n$ so that~(\ref{eq:dis_code_c}) is satisfied. From~(\ref{eq:cons_ach_dbound1}) and (\ref{eq:code-size_c}) we conclude that a sequence of $(T_n,{R}+\epsilon,{D}+\epsilon)$ code exists with $T_n=n\Delta$ and $T_n\to\infty$ as $n\to\infty$.

\paragraph*{Converse}
We will prove the converse when feedforward is present.
For the given $(T,R+\epsilon,D+\epsilon)$ code with feedforward, let $M$ denote the encoder's output. Since $I(M;Y_0^T)<\infty$, we conclude from Theorem~\ref{thm:MI} that there exists 
a process $\Gamma_0^T$ such that $\Gamma_0^T$ is the $(\F_t=\sigma(M,Y_0^t):\tin)$ intensity of $Y_0^T$ and 
\begin{equation}
I(M;Y_0^T)=\E\left[\int_0^T \phi(\Gamma_t)\,dt\right]-T\phi(\lambda).
\end{equation}
We also have 
\begin{align*}
\frac{1}{T}I(M;{Y}^T_0)=\frac{1}{T}H(M) \le \frac{1}{T}\log\left(\lceil \exp((R+\epsilon)T) \rceil \right) \le R + \epsilon+\frac{1}{T}.
\end{align*}
This gives 
\begin{align*}
R\ge \frac{1}{T}\E\left[\int_0^T \phi(\Gamma_t)\,dt\right]-\phi(\lambda)-\epsilon-\frac{1}{T}.
\end{align*}
Let $\hat{Y}_0^T$ denote the decoder's output. 
The distortion constraint $D$ satisfies
\begin{align}
D&\ge \frac{1}{T}\E\left[d(\hat{Y}_0^T,Y_0^T) \right]-\epsilon\nn
&=\frac{1}{T}\E\left[\int_0^T \hat{Y}_t\,dt-\log(\hat{Y}_t)\,dY_t\right]-\epsilon\nn
&\overset{(a)}{=}\frac{1}{T}\E\left[\int_0^T \hat{Y}_t-\Gamma_t\log(\hat{Y}_t)\,\,dt\right]-\epsilon\nn
&\ge \frac{1}{T}\E\left[\int_0^T \inf_{v\in\A } v-\Gamma_t\log(v)\,dt\right]-\epsilon\nn
&\overset{(b)}{=}\frac{1}{T}\E\left[\int_0^T \Psi_{\A}(\Gamma_t)\,dt\right]-\epsilon,
\end{align}
where, for (a) we have used Lemma~\ref{Le:hat_Yt}, and \\*
for (b), we have used the definition of $\Psi_\A$.
Defining $S$  to be uniformly distributed on $[0,T]$, and independent of all other random variables we have
\begin{align}
R&\ge \E\left[\phi(\Gamma_S)\right]-\phi(\lambda)-\epsilon-\frac{1}{T} \label{eq:CFC_rdconv_r-i_1}\\
D&\ge \E\left[\Psi_{\A}(\Gamma_S)\right]-\epsilon, \label{eq:CFC_rdconv_r-i_2}
\end{align}

Now we use Carath\'{e}odory's  theorem~\cite[Theorem 17.1]{Rockafellar'97}. There exist  non-negative $[\eta_k]_{k=1}^4$ and $[\alpha_k]_{k=1}^4$, such that $\sum_{k=1}^4 \alpha_k =1$ and 
\begin{align}
\E\left[\phi({\Gamma}_{S})\right]&=\sum_{k=1}^4 \alpha_k\phi(\eta_k),\\
\E\left[\Psi_{\A}({\Gamma}_{S})\right]&=\sum_{k=1}^4 \alpha_k\Psi_\A(\eta_k),\\
\E\left[{\Gamma}_{S}\right]&=\sum_{k=1}^4 \alpha_k\eta_k=\lambda,
\end{align}
where in the last line we have used the fact that since ${\Gamma}_0^{T}$ is the $(\sigma(M,{Y}_{0}^T):\tin)$-intensity of ${Y}_0^{T}$,
$\E\left[\int_0^T{\Gamma}_{t}\,dt\right]=\E[{Y}_T]=T\lambda$.
Now define 
$$\beta_k\de\frac{\alpha_k\eta_k}{\lambda}.$$ 
We note that $\beta_k=0$ if $\alpha_k=0$, and $\sum_{k=1}^4\beta_k=1$. Substituting the above definitions in~(\ref{eq:CFC_rdconv_r-i_1})-(\ref{eq:CFC_rdconv_r-i_2}), we obtain
 \begin{align}
 R &\ge\left(\sum_{k=1}^4 \alpha_k\eta_k\log(\eta_k)-\lambda\log(\lambda)\right)-\epsilon-\frac{1}{T}\nn
 &=\lambda\left( \sum_{k=1}^4 \beta_k\log\left(\frac{\beta_k\lambda}{\alpha_k}\right)\1\{\alpha_k>0\}-\log(\lambda)\right)-\epsilon-\frac{1}{T}\nn
 &=\lambda\sum_{k=1}^4 \beta_k\log\left(\frac{\beta_k}{\alpha_k}\right)-\epsilon-\frac{1}{T}.
 \end{align}
Likewise,
\begin{align*}
D\ge \sum_{k=1}^4 \alpha_k\Psi_\A\left(\frac{\lambda\beta_k}{\alpha_k}\right)-\epsilon.
\end{align*}
 
 Since $\epsilon$ is arbitrary and $T$ can be made arbitrarily large, we obtain the rate-distortion region in the statement of the theorem. 
\end{IEEEproof}

If we do not place any restrictions on $\A$, i.e., if $\A$ is the set of all non-negative reals, then we obtain the functional-covering distortion.
\begin{Corollary}[Functional Covering of Poisson Processes]
The rate-distortion function for functional-covering distortion is given by $R_{\text{FC}}(D)=(\lambda-\lambda\log(\lambda)-D)^+$.
\begin{IEEEproof}
For the functional-covering distortion, $\A$ is the set of non-negative reals. Hence
\begin{align*}
\Psi_\A(u)=\inf_{v\ge 0} v-u\log (v)=u-u\log(u).
\end{align*}
For any achievable $(R,D)$ we have
\begin{align}
R &\ge \lambda\sum_{k=1}^4 \beta_k\log\left(\frac{\beta_k}{\alpha_k} \right),
\label{eq:R_FCD}
\end{align}
and
\begin{align*}
D &\ge \sum_{k=1}^4  \alpha_k \Psi_\A\left(\frac{\lambda\beta_k}{\alpha_k}\right)\\
&=\sum_{k=1}^4 \alpha_k \left(\frac{\lambda\beta_k}{\alpha_k}-\frac{\lambda\beta_k}{\alpha_k}\log\left(\frac{\lambda\beta_k}{\alpha_k} \right)\right)\\
&=\lambda-\lambda\log(\lambda)-\lambda\sum_{k=1}^4 \beta_k\log\left(\frac{\beta_k}{\alpha_k} \right).
\end{align*}
Hence
\begin{align*}
R+D\ge \lambda-\lambda\log(\lambda),
\end{align*}
and this is achieved by $[\alpha_k]_{k=1}^4$ and $[\beta_k]_{k=1}^4$ that yield equality in~(\ref{eq:R_FCD}).
\end{IEEEproof}

\end{Corollary}
If take $\A=\{0,1\}$, then we recover the covering distortion in~\cite[Theorem 1]{Lapidoth2015}.
\begin{Corollary}[Covering Distortion~\cite{Lapidoth2015}]
The rate-distortion function for the covering distortion is given by $R_{\text{C}}(D)=(-\lambda\log(D))^+$.
\begin{IEEEproof}
For the covering distortion, $\A=\{0,1\}$. Hence
\begin{align*}
\Psi_\A(u)=\inf_{v\in\{0,1\}} v-u\log (v)=\1\{u>0\}.
\end{align*}
Suppose $(R,D)$ is in $\mathfrak{RD}$. Then
\begin{align*}
D &\ge \sum_{k=1}^4  \alpha_k \Psi_\A\left(\frac{\lambda\beta_k}{\alpha_k}\right)\\
&=\sum_{k=1}^4 \alpha_k \1\{\beta_k>0\}\\
&=\sum_{k\in \mathcal{B}}\alpha_k,
\end{align*}
where we have defined $\mathcal{B}=\{k:\beta_k>0\}$. Similarly,
\begin{align*}
R &\ge \lambda\sum_{k=1}^4 \beta_k\log\left(\frac{\beta_k}{\alpha_k} \right)\\
&= \lambda\sum_{k\in\mathcal{B}} \beta_k\log\left(\frac{\beta_k}{\alpha_k} \right)\\
& \overset{(a)}{\ge} \lambda\left(\sum_{k\in\mathcal{B}}\beta_k\right)\log\left(\frac{\sum_{k\in\mathcal{B}}\beta_k}{\sum_{k\in\mathcal{B}}\alpha_k} \right)\\
&= \lambda \log\left(\frac{1}{\sum_{k\in\mathcal{B}}\alpha_k} \right)\\
&\ge \left(-\lambda \log(D) \right)^+,
\end{align*}
where, (a) is due to the log-sum inequality, which can be achieved by setting $\alpha_1=\min(1,D)$, $\alpha_2=1-\alpha_1$, $\beta_1=1$, $\beta_2=0$.
\end{IEEEproof}
\end{Corollary}

\begin{Remark}
As in the general case in Theorem~\ref{thm:R-D} (see Remark~\ref{rem:thmR_D}), the reconstruction $\hat{y}_0^T$ (assuming it is bounded) can be used to define a probability measure on the input space  $(\mathcal{N}_0^T,\Fk^N)$ via 
\[
\frac{dP_{\hat{y}_0^T}}{dP_0}(y_0^T) = \exp\left(\int_0^T\log( \hat{y}_t)\,dy_t-(\hat{y}_t-1)\,dt \right),
\]
where $P_0$ is the measure under which $Y_0^T$ is a Poisson process with unit rate. Moreover, in absence of feedforward, $\hat{y}_0^T$ is deterministic (it depends only  on the encoder's output). Thus the input point-process $Y_0^T$ is a non-homogeneous Poisson processes with rate $\hat{y}_0^T$ under $P_{\hat{y}_0^T}$. As in the general case, the functional-covering distortion is related to the above Radon-Nikodym derivative via 
\[
d(\hat{y}_0^T,y_0^T) = -\log\left(\frac{dP_{\hat{y}_0^T}}{dP_{0}} (y_0^T)\right)+T
\]
\eor
\end{Remark}

\section{The Poisson CEO Problem}
\begin{center}
\begin{figure}
    \begin{center}
\includegraphics[scale=0.45]{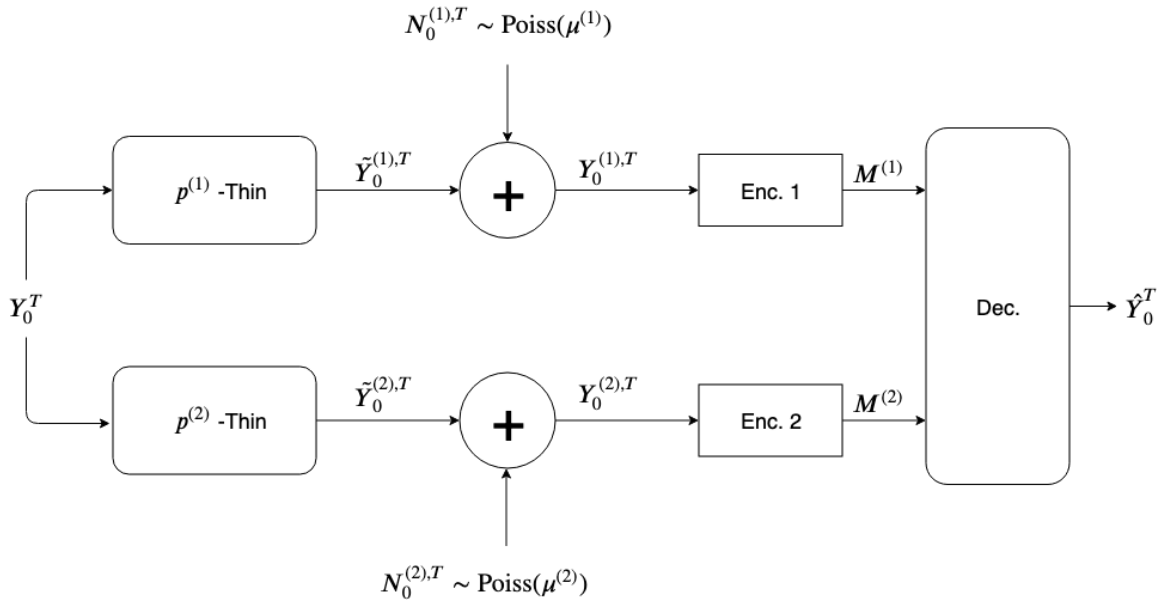}
    \end{center}
\caption{Poisson CEO Problem.}
\label{fig:Poisson_CEO}
\end{figure}
\end{center}

We now consider the distributed problem shown Figure~\ref{fig:Poisson_CEO}. Our goal is to compress $Y_0^T$, which is a Poisson process with rate $\lambda>0$. Each of the two encoders observes a degraded version of $Y_0^T$, denoted by $Y_{0}^{(i), T}$, $i\in\{1,2\}$.  $Y_0^T$ is first $p^{(i)}$-thinned to obtain $\tilde{Y}_{0}^{(i), T}$,  and then  an independent Poisson process $N_{0}^{(i), T}$ with rate $\mu^{(i)}$ is added to $\tilde{Y}_{0}^{(i), T}$ to obtain $Y_{0}^{(i), T}$.

Recall that $\hat{\mathcal{Y}}_0^T$ is the set of all non-negative functions $\hat{y}_0^T$ which are left-continuous with right-limits, and
$$
d(\hat{y}_0^T,y_0^T)=\int_0^T \hat{y}_t\,dt-\log(\hat{y}_t)\,dy_t.
$$ 
\begin{Def}
A $(T,R^{(1)},R^{(2)},D)$ code for the Poisson CEO problem consists of  \emph{encoders} $f^{(1)}$ and $f^{(2)}$,
\begin{align*}
f^{(1)}:\mathcal{N}_0^T \rightarrow \{1,\dots,\lceil\exp(R^{(1)}T)\rceil \}\\
f^{(2)}:\mathcal{N}_0^T \rightarrow \{1,\dots,\lceil\exp(R^{(2)}T)\rceil \},
\end{align*}
and a \emph{decoder} $g$,
\begin{align*}
g:\{1,\dots,\lceil\exp(R^{(1)}T)\rceil\}\times\{1,\dots,\lceil\exp(R^{(2)}T)\rceil\} \rightarrow \hat{\mathcal{Y}}_0^T,
\end{align*}
satisfying
\begin{align*}
\E\left[\int_0^T \hat{Y}_t\,dt\right]<\infty,
\end{align*}
and the distortion constraint 
\begin{align*}
\frac{1}{T}\E\left[d(\hat{Y}_0^T,Y_0^T)  \right] \le D.
\end{align*}
\end{Def}
\begin{Def}
A {rate-distortion vector} $(R^{(1)},R^{(2)},D)$ is said to be \emph{achievable} for the Poisson CEO problem if for any $\epsilon>0$, there exists a sequence $(T_n,R^{(1)}+\epsilon,R^{(2)}+\epsilon,D+\epsilon)$ codes  $T_n\to\infty$ .
\end{Def}
\begin{Def}
The \emph{rate-distortion region} for the Poisson CEO problem $\mathfrak{RD}^{\mathcal{P}}$ is the intersection of all achievable rate-distortion vectors $(R^{(1)},R^{(2)},D)$.
\end{Def}


The rate-distortion region for the Poisson CEO problem with feedforward, denoted by $\mathfrak{RD}^{\mathcal{P}}_F$, is defined analogously. 
\begin{Theorem}
\label{thm:Poisson_CEO}
The rate-distortion region for the Poisson CEO problem is given by
$$\mathfrak{RD}^{\mathcal{P}}=\mathfrak{RD}^{\mathcal{P}}_F=\mathfrak{RD},$$ where $\mathfrak{RD}$ is the convex hull of union of sets of rate-distortion vectors  $(R^{(1)},R^{(2)},D)$ such that 
\begin{gather*}
R^{(1)} \ge \left((1-p^{(1)})\lambda+\mu^{(1)}\right) \sum_{k=1}^4 \beta^{(1)}_{k}\log\left(\frac{\beta^{(1)}_{k}}{\alpha^{(1)}_{k}}\right),\\
R^{(2)} \ge \left((1-p^{(2)})\lambda+\mu^{(2)}\right)\sum_{k=1}^4 \beta^{(2)}_{k}\log\left(\frac{\beta^{(2)}_{k}}{\alpha^{(2)}_{k}}\right),\\
D \ge \lambda-\phi(\lambda)-\lambda\left(\sum_{k=1}^4 \gamma^{(1)}_{k}\log\left(\frac{\gamma^{(1)}_{k}}{\alpha^{(1)}_{k}}\right)+\sum_{k=1}^4\gamma^{(2)}_{k}\log\left(\frac{\gamma^{(2)}_{k}}{\alpha^{(2)}_{k}}\right)\right)
\end{gather*}
for some probability vectors $[\alpha^{(i)}_{k}]_{k=1}^4$, $[\beta^{(i)}_{k}]_{k=1}^4$, and $[\gamma^{(i)}_{k}]_{k=1}^4$, where for $\kin$ and $i\in\{1,2\}$
\begin{align*}
&\begin{rcases}
\gamma^{(i)}_{k}=p^{(i)}\alpha^{(i)}_{k}+(1-p^{(i)})\beta^{(i)}_{k}\\
\alpha^{(i)}_{k}=0 \Rightarrow \beta^{(i)}_{k}=0
\end{rcases}
\quad\text{if } p^{(i)}<1,\\
&\alpha^{(i)}_{k}=\beta^{(i)}_{k}=\gamma^{(i)}_{k} \quad\qquad\qquad\quad\quad\,\,\,\text{if } p^{(i)}=1.
\end{align*}
\end{Theorem}

\begin{IEEEproof}
    Please see the supplementary material.
\end{IEEEproof}

\begin{Remark}
\label{rem:BT_VNC}
    Note that there is no sum-rate constraint in the rate-distortion region of the above theorem. This occurs due to the sparsity of points in a Poisson process. After discretizing a Poisson process with rate $\lambda$, the expected number of ones in the resulting binary process is roughly $\lambda T$, and the remaining $T/\Delta-\lambda T$ bits are zeroes. When such a sparse binary process is sent via two independent parallel channels as in~(\ref{EQ:test-ch1})-(\ref{EQ:test-ch2}), the  resulting output processes are almost independent.  
    This implies that the encoders do not need to bin their messages in the achievability argument.  
\end{Remark}

\begin{Corollary}[Poisson CEO Problem without Thinning]
If $p^{(1)}=p^{(2)}=0$, then the rate-distortion region in Theorem~\ref{thm:Poisson_CEO} takes a simple form
$$
\frac{\lambda}{\lambda+\mu^{(1)}}R^{(1)}+\frac{\lambda}{\lambda+\mu^{(2)}}R^{(2)}+D\ge \lambda-\phi(\lambda).
$$
\end{Corollary}

\begin{Corollary}[Remote Poisson Source]
\label{cor:RPS}
Consider a scenario where an encoder wishes to compress a Poisson process with rate $\lambda>0$, but observes a degraded version of it, where the points are first erased with independent probability $1-p$ and then an independent Poisson process with rate $\mu$ is added to it.
Then the rate-distortion region $(R,D)$  is the convex hull of the union of all rate-distortion vectors satisfying
\begin{gather*}
R \ge \left((1-p)\lambda+\mu\right) \sum_{k=1}^4 \beta_{k}\log\left(\frac{\beta_{k}}{\alpha_{k}}\right),\\
D \ge \lambda-\phi(\lambda)-\lambda\cdot\sum_{k=1}^4 \gamma_{k}\log\left(\frac{\gamma_{k}}{\alpha_{k}}\right),
\end{gather*}
for some probability vectors $[\alpha_{k}]_{k=1}^4$, $[\beta_{k}]_{k=1}^4$, and $[\gamma_{k}]_{k=1}^4$, where for $k\in\{1,2,3,4\}$
\begin{align*}
\gamma_{k}=p\alpha_{k}+(1-p)\beta_{k},\quad
\alpha_{k}=0 \Rightarrow \beta_{k}=0.
\end{align*}
\end{Corollary} 

\bibliographystyle{IEEEtran}
\bibliography{IEEEabrv,Poisson}

\newpage

\appendix
\begin{IEEEproof}[Proof of Lemma~\ref{lem:abs_cont}]
The first part of the lemma is due to~\cite[T12 Theorem, Chapter VI, p. 187]{Bremaud}. 
To prove the second part we note that $\E_{P^{Y_0^T}}[\int_{0}^T|\phi(\Gamma_t)|\,dt]<\infty$ implies $\E_{P^{Y_0^T}}[\int_{0}^T\Gamma_t\,dt]<\infty$, which in turn gives $$\int_{0}^T(1-\sqrt{\Gamma_t}))^2\le \int_{0}^T(\Gamma_t+1)<\infty,$$ ${P^{Y_0^T}}$-a.s. Thus applying~\cite[Theorem 19.7, p. 343]{Liptser}, we conclude that $P^{Y_0^T} \ll P_0^{Y_0^T}$. Hence, from the first part of the lemma
\begin{align*}
\frac{dP^{Y_0^T}}{dP_0^{Y_0^T}}=\exp\left(\int_0^T\log(\Lambda_t)\,dY_t-\Lambda_t+1\,dt\right),
\end{align*}
where the uniqueness of intensity~\cite[T12 Theorem, Chapter II, p. 31]{Bremaud} gives us  
\begin{align*}
\E_{P^{Y_0^T}}\left[\int_0^T|\Gamma_t-\Lambda_t|\,dt\right]=0,\quad \E_{P^{Y_0^T}}\left[\int_0^T\1\{\Gamma_t\ne\Lambda_t\}\,dY_t\right]=0.
\end{align*}

Since 
$$\E_{P^{Y_0^T}}\left[\int_{0}^T|\phi(\Gamma_t)|\,dt\right]<\infty,$$ 
we have 
$$\E_{P^{Y_0^T}}\left[\int_{0}^T(\log(\Gamma_t))^+\,dY_t\right]=\E_{P^{Y_0^T}}\left[\int_{0}^T(\log(\Gamma_t))^+\Gamma_t\,dt\right]=\E_{P^{Y_0^T}}\left[\int_{0}^T(\phi(\Gamma_t))^+\,dt\right]<\infty,$$ 
and 
$$\E_{P^{Y_0^T}}\left[\int_{0}^T(\log(\Gamma_t))^-\,dY_t\right]=\E_{P^{Y_0^T}}\left[\int_{0}^T(\log(\Gamma_t))^-\Gamma_t\,dt\right]=\E_{P^{Y_0^T}}\left[\int_{0}^T(\phi(\Gamma_t))^-\,dt\right]<\infty.$$ 
Hence $$\E_{P^{Y_0^T}}\left[\int_{0}^T\log(\Gamma_t)\,dY_t\right]=\E_{P^{Y_0^T}}\left[\int_{0}^T\phi(\Gamma_t)\,dt\right]<\infty.$$

Finally,
\begin{align*}
\E_{P^{Y_0^T}}\left[\log\left(\frac{dP^{Y_0^T}}{dP_0^{Y_0^T}}\right)\right]&=\E_{P^{Y_0^T}}\left[\int_0^T\log(\Lambda_t)\,dY_t+\Lambda_t-1\,dt\right]\\
&\aeq\E_{P^{Y_0^T}}\left[\int_0^T\log(\Gamma_t)\,dY_t+\Gamma_t-1\,dt\right]\\
&=\E_{P^{Y_0^T}}\left[\int_0^T\log(\Gamma_t)\,dY_t\right]-\E\left[\int_0^T\Gamma_t-1\,dt\right]\\
&=\E_{P^{Y_0^T}}\left[\int_0^T\phi(\Gamma_t)\,dt\right]-\E\left[\int_0^T\Gamma_t-1\,dt\right]\\
&=\E_{P^{Y_0^T}}\left[\int_0^T\phi(\Gamma_t)-\Gamma_t+1\,dt\right].
\end{align*}
Here, for (a) we have used the uniqueness of the intensity and in the remaining equalities, we have used the finiteness of the expectations $\left[\int_0^T\phi(\Gamma_t)\,dt\right]$, $\E\left[\int_0^T\Gamma_t\,dt\right]$.
\end{IEEEproof}

\begin{IEEEproof}[Proof of Lemma~\ref{Le:int_eqn}]
Recall that $L_0^T$ can be written as
\begin{align*}
L_t=\exp\left(\int_{0}^t\log (\Gamma_s)\,dY_s+(1-\Gamma_s)\,ds\right),\quad \tin.
\end{align*}
We note that for $\tin$ $L_t$ satisfies
\begin{align}
L_t=\begin{cases}
L_{t-} & \text{if }Y_t-Y_{t-} =0,\\
\Gamma_tL_{t-} & \text{if }Y_t-Y_{t-} =1.
\end{cases}
\label{eq:app1}
\end{align}
Let $C_0^T$ be a non-negative $(\G_t:\tin)$-predictable process. Then
\begin{align*}
\E\left[\int_0^T C_t\,dY_t\right]&\overset{(a)}{=}\E_{\tilde{P}^{M,Y_0^T}}\left[L_T\int_0^T  C_t\,dY_t\right]\\
&\overset{(b)}{=}\E_{\tilde{P}^{M,Y_0^T}}\left[\int_0^T  L_tC_t\,dY_t\right]\\
&\overset{(c)}{=}\E_{\tilde{P}^{M,Y_0^T}}\left[\int_0^T  \Gamma_tL_{t-}C_t\,dY_t\right]\\
&\overset{(d)}{=}\E_{\tilde{P}^{M,Y_0^T}}\left[\int_0^T  \Gamma_tL_{t-}C_t\,dt\right]\\
&\overset{(e)}{=}\E_{\tilde{P}^{M,Y_0^T}}\left[\int_0^T  \Gamma_tL_{t}C_t\,dt\right]\\
&\overset{(f)}{=}\E_{\tilde{P}^{M,Y_0^T}}\left[L_T\int_0^T  \Gamma_tC_t\,dt\right]\\
&\overset{(g)}{=}\E_{}\left[\int_0^T  \Gamma_tC_t\,dt\right],
\end{align*}
where, (a) follows since $L_T$ is the Radon-Nikodym derivative $\frac{d{P}^{M,Y_0^T}}{d\tilde{P}^{M,Y_0^T}}$,\\*
(b) follows due to~\cite[T19 Theorem, Appendix A2, p. 302]{Bremaud},\\*
(c) follows due to~(\ref{eq:app1}),\\*
(d) follows since the $(\tilde{P}^{M,Y_0^T},\G_t:\tin)$-intensity of $Y_0^T$ is 1, and $L_{t-}$ being a left-continuous adapted process is  $(\G_t:\tin)$-predictable,\\*
(e) follows since the Lebesgue measure of the set $\{t:\tin,L_{t-}\neq L_t\}$ is zero due to~(\ref{eq:app1}),\\*
(f) again follows again due to~\cite[T19 Theorem, Appendix A2, p. 302]{Bremaud}, and \\*
(g) again follows  since $L_T$ is the Radon-Nikodym derivative $\frac{d{P}^{M,Y_0^T}}{d\tilde{P}^{M,Y_0^T}}$.
\end{IEEEproof}

\begin{IEEEproof}[Proof of Lemma~\ref{Le:phi_fun}]
We will first show that $$\E\left[\int_{0}^{T}\left(\log (\Gamma_t)\right)^-\,dY_t\right]=\E\left[\int_{0}^{T}\left(\log (\Gamma_t)\right)^-\Gamma_t\,dt\right]<\infty.$$
Define $\Gamma^{1+}_t\de\max(\Gamma_t,1)$ and $\Gamma^{1-}_t\de\min(\Gamma_t,1)$. We note that $\Gamma_t\le\Gamma^{1+}_t\le\Gamma_t+1$ and $\Gamma_t=\Gamma^{1+}_t\Gamma^{1-}_t$. Define the process $\mu_0^T$ as
\begin{align*}
\mu_t\de\frac{\Gamma^{1+}_t}{\Gamma_t}\1\{\Gamma_t>0\}, \quad \tin.
\end{align*}
Then $\mu_0^T$ is a non-negative $(\G_t:\tin)$-predictable process and $$\int_0^T \mu_t \Gamma_t\,dt=\int_0^T \Gamma^{1+}_t\1\{\Gamma_t>0\}\,dt\le\int_0^T  (\Gamma_t+1)\,dt<\infty$$ $P$-a.s. since 
$\E[\int_0^T\Gamma_t\,dt]<\infty$. Hence the process $\hat{L}_0^T$ defined as
\begin{align*}
\hat{L}_t \de \exp \left(\int_0^T \log(\mu_t)\,dY_t+(1-\mu_t)\Gamma_t\,dt  \right), \quad \tin
\end{align*}
is a $(P,G_t:\tin)$ non-negative super-martingale~\cite[T2 Theorem, Chapter VI, p. 165]{Bremaud}. Hence the following chain of inequalities hold
\begin{align}
\E\left[\log (\hat{L}_T)\right]
&\overset{(a)}{\le} \log \left(\E[\hat{L}_T]\right)\nn
&\overset{(b)}{\le} \log \left(\E[\hat{L}_0]\right)\nn
&=0. \label{eq:sumab}
\end{align}
Here, for (a) we have used the fact that  since $L_0^T$ is a super-martingale, $L_T$ is integrable, and then Jensen's inequality and \\*
for (b), we have used the fact that $\hat{L}_0^T$ is a super-martingale, hence $\E[\hat{L}_T]\le\E[\hat{L}_0]$.\\*
Let $\tau_k$ denote the $k$th arrival instant of the process $Y_0^T$, i.e.,
\begin{align*}
\tau_k = \inf\{\tin:Y_t=k\},
\end{align*}
where the infimum of the null set is taken as $\infty$. Then if $\tau_k\le T$,  $\Gamma_{\tau_k}>0$ $P$-a.s.~\cite[T12 Theorem, Chapter II, p. 31]{Bremaud}. Hence for $\tau_k\le T$,
\begin{align*}
\log(\mu_{\tau_k})=\log(\Gamma^{1+}_{\tau_k})-\log(\Gamma_{\tau_k})=-\log(\Gamma^{1-}_{\tau_k})=\left(\log(\Gamma^{}_{\tau_k})\right)^- \quad P-\text{a.s.},
\end{align*}
Thus we can write
\begin{align*}
\log (\hat{L}_T)=\int_0^T \left(\log (\Gamma_t)\right)^-\,dY_t+\int_0^T(\Gamma_t-\Gamma^{1+}_t)\1\{\Gamma_t>0\}\,dt.
\end{align*}
Using~(\ref{eq:sumab}) we obtain
\begin{align*}
\E\left[\int_0^T\left(\log (\Gamma_t)\right)^-\,dY_t+\int_0^T(\Gamma_t-\Gamma^{1+}_t)\1\{\Gamma_t>0\}\,dt\right] =\E[\log(\hat{L_T})]\le 0.
\end{align*}
We note that $\int_0^T \left(\log (\Gamma_t)\right)^-\,dY_t$ is a non-negative random variable, and  $$\left|\E\left[\int_0^T(\Gamma_t-\Gamma^{1+}_t)\1\{\Gamma_t>0\}\,dt\right]\right|\le \E\left[\int_0^T(\Gamma_t+\Gamma^{1+}_t)\,dt\right]\le\E\left[\int_0^T(2\Gamma_t+1)\,dt\right]<\infty.$$ Hence we can split the expectation to get
\begin{align*}
\E\left[\int_0^T\left(\log (\Gamma_t)\right)^-\,dY_t\right]+\E\left[\int_0^T(\Gamma_t-\Gamma^{1+}_t)\1\{\Gamma_t>0\}\,dt\right]\le
0,
\end{align*}
which gives
\begin{align}
\E\left[\int_0^T \left(\log (\Gamma_t)\right)^-\,dY_t\right]\le-\E\left[\int_0^T(\Gamma_t-\Gamma^{1+}_t)\1\{\Gamma_t>0\}\,dt\right]<\infty.
\label{eq:mi-proof-7}
\end{align}
Hence
\begin{align}
\E\left[\int_{0}^{T}\log (\Gamma_t)\,dY_t\right]&=\E\left[\int_{0}^{T}\left(\log (\Gamma_t)\right)^+\,dY_t\right]-\E\left[\int_{0}^{T}\left(\log (\Gamma_t)\right)^-\,dY_t\right]\nn
&=\E\left[\int_{0}^{T}(\log (\Gamma_t))^+\Gamma_t\,dt\right]-\E\left[\int_{0}^{T}\left(\log (\Gamma_t)\right)^-\Gamma_t\,dt\right]\nn
&=\E\left[\int_{0}^{T}\phi (\Gamma_t)\,dt\right].
\label{eq:mi-proof-1}
\end{align}
\end{IEEEproof}

\begin{IEEEproof}[Proof of Lemma~\ref{lem:martingale_int}]
Suppose that $\Gamma_0^T$ is the $(\F_t:\tin)$-intensity of  $Y_0^T$. Then applying~\cite[T8 Theorem, Chapter II, p. 27]{Bremaud} with $X_s=1$ proves $M_0^T$ is  a $(\F_t:\tin)$-martingale.
Now suppose that $M_0^T$ is a $(\F_t:\tin)$-martingale. Consider a simple $(\mathcal{F}_t:\tin)$-predictable process $C_0^T$ of the  form 
\begin{align*}
C_t=\1\{\EE\}\1\{u<t\le v \le T\}\quad\EE\in\mathcal{F}_u.
\end{align*}
Then
\begin{align}
\E\left[\int_0^T C_s \, d\Y_{s}\right]&=\E\left[\1\{\EE\}(Y_v-Y_u)\right]\nn
&=\E\left[\1\{\EE\}\E[(Y_v-Y_u)|\mathcal{F}_u]\right]\nn
&\overset{(a)}{=}\E\left[\1\{\EE\}\E\left[\int_u^v \Gamma_s\,ds\middle|\mathcal{F}_u\right]\right]\nn
&=\E\left[\int_0^T C_s \Gamma_s\, d{s}\right],
\label{eq:thin_int_1}
\end{align}
where for (a) we have used the martingale property of $M_0^T$.
Thus by the monotone class theorem, for all bounded $(\mathcal{F}_t:\tin)$-predictable processes $C_0^T$, (\ref{eq:thin_int_1}) holds (see~\cite[App.~A1, Theorem T5, p.~264]{Bremaud}). Then by applying the monotone convergence theorem, we can show that (\ref{eq:thin_int_1}) holds for all non-negative $(\mathcal{F}_t:\tin)$-predictable processes as well, so  that $\Gamma_0^T$ is the $(\F_t:\tin)$-intensity of  $Y_0^T$.
\end{IEEEproof}

\begin{IEEEproof}[Proof of Lemma~\ref{lem:int_his}]
There exists a $(\G_t:\tin)$-predictable process $\Pi_0^T$ such that $P$-a.s.
 $\Pi_t=\E[{\Lambda}_t|\G_{t-}]$, $\tin$~\cite[Chapter 6, Theorem 43, p. 103]{Dellacherie}. We will show that $\Pi_0^T$   is the $(\G_t:\tin)$-intensity of ${N}_0^T$. Let ${D}_0^T$ be a non-negative $(\G_t:\tin)$-predictable process. As $\G_t\subseteq\F_t$, it is also $(\F_t:\tin)$-predictable.
Thus 
\begin{align}
\E\left[\int_0^{T} D_s \, d{\N}_{s}\right]=\E\left[\int_0^{T} D_s {\Lambda}_s \, ds\right].
\label{EQ:intensity2}  
\end{align}
Hence
\begin{align*}
\E\left[\int_0^{T} D_s \Pi_s \, ds\right]&=\E\left[\int_0^{T} D_s \E[{\Lambda}_s|\G_{s-}] \, ds\right]\nn
&\stackrel{(a)}{=}\E\left[\int_0^{T} \E[D_s {\Lambda}_s|\G_{s-}] \, ds\right]\nn
&=\E\left[\int_0^{T} D_s {\Lambda}_s \, ds\right]\nn
&\stackrel{(b)}{=}\E\left[\int_0^{T} D_s \, d{\N}_{s}\right].
\end{align*}
Here, (a) is due to the fact that $D_s$ is $\G_{s-}$ measurable \cite[Exercise E10, Chapter I, p. 9]{Bremaud}, and\\*
(b) is due to (\ref{EQ:intensity2}).

Hence the $\left(\G_t:\tin\right)$-intensity of ${N}_0^{{T}}$ is $\Pi_0^{{T}}$. 
\end{IEEEproof}

\begin{IEEEproof}[Proof of Lemma~\ref{lem:add_int}]
We first note that since $Y_0^T$ and $N_0^T$ are independent,  trajectories of 
$Z_0^T$ are a.s. in $\mathcal{N}_0^T$. 
The $(\tF_t\de\sigma(M,Y_0^t,N_0^t):\tin)$-intensities of $Y_0^T$ and $N_0^T$ are $\Gamma_0^T$ and  $\Pi_0^T$ respectively~\cite[E5 Exercise, Chapter II, p. 28]{Bremaud}. 
Then for  a non-negative $(\tF_t:\tin)$-predictable process $C_0^T$:
\begin{align*}
\E\left[\int_0^T C_s \, d\Z_{s}\right]&=\E\left[\int_0^T C_s \, d\Y_{s}\right]+\E\left[\int_0^T C_s \, d\N_{s}\right]\\
&=\E\left[\int_0^T C_s \Gamma_s \, d{s}\right]+\E\left[\int_0^T C_s \Pi_s \, ds\right]\\
&=\E\left[\int_0^T C_s (\Gamma_s+\Pi_s) \, d{s}\right].
\end{align*}
Hence the $(\tF_t:\tin)$-intensity of $Z_0^T$ is $(\Gamma_t+\Pi_t:\tin)$. Since $\F_t\subseteq\tF_t$ the statement of the lemma follows from an application of Lemma~\ref{lem:int_his}.
\end{IEEEproof}

\begin{IEEEproof}[Proof of Lemma~\ref{lem:thin_int}]
Let $\mathcal{H}_t\de\sigma(M,Y_0^t,Z_0^t)$. We  note that the $(\mathcal{H}_t:\tin)$-intensity of $Y_0^T$ is $\Gamma_0^T$. 
Now we will compute the $(\mathcal{H}_t:\tin)$-intensity of $Z_0^T$. 
Let $(\chi_i:i\in\{1,\dots\})$ denote the sequence of independent and identically distributed Bernoulli random variables which indicate if a particular point in point process $\Y_0^T$ is erased or not. In particular, if $\chi_j=1$, then the $j$th point in $Y_0^T$ is retained, so that $\E[\chi_j]=1-p$. Then for $0\le u < v \le T$
\begin{align*}
Z_v-Z_u=\sum_{k=Y_u+1}^{Y_v}\chi_k=\sum_{k=1}^{\infty}\chi_k\1\{Y_u< k\le Y_v\}.
\end{align*} 
Using the monotone convergence theorem for the conditional expectation, 
\begin{align*}
\E\left[(Z_v-Z_u)|\mathcal{H}_u\right]&=\sum_{k=1}^{\infty}\E[\chi_k\1\{Y_u< k\le Y_v\}|\mathcal{H}_u]\\
&\overset{(a)}{=}\sum_{k=1}^{\infty}\E[\chi_k|\mathcal{H}_u]\E[\1\{Y_u< k\le Y_v\}|\mathcal{H}_u]\\
&\overset{(b)}{=}(1-p)\E[(Y_v-Y_u)|\mathcal{H}_u]\\
&\overset{(c)}{=} (1-p)\E\left[\int_u^v \Gamma_s\,ds\middle|\mathcal{H}_u\right],
\end{align*}
where, for (a) we have used the fact that given $\mathcal{H}_u$, $\chi_k$ is independent of $Y_0^T$,\\*
for (b), we use note that $\E[\chi_k|\mathcal{H}_u]=\chi_k\1\{k\le Y_u\}+(1-p)\1\{k> Y_u\}$, and \\*
for (c), we have used the martingale property of $M_0^T$.\\*
Then
$$
\tilde{M}_t\de Z_t-\int_0^t (1-p)\Gamma_s\,ds \quad\tin
$$
is a $(\mathcal{H}_t:\tin)$-martingale. Hence from  Lemma~\ref{lem:martingale_int}, the $(\mathcal{H}_t:\tin)$-intensity of $Z_0^T$ is $((1-p)\Gamma_t:\tin)$. An application of Lemma~\ref{lem:int_his} the proves the statement of the lemma.
\end{IEEEproof}

\begin{IEEEproof}[Proof of Lemma~\ref{Le:hat_Yt}]
We will require the following inequality 
\begin{align}
u\log(v)\le \phi(u)-u+v, \quad 0\le u,v < \infty.
\label{eq:ineq}
\end{align}
The inequality can be verified to be true if either or both  $u$, $v$ are zero. If $u,v>0$, the inequality follows from $\log(u/v)\ge(1-v/u) $.

Defining $\hat{Y}^{1+}_t\de\max(1,\hat{Y}_t)$, we note that $\hat{Y}^{1+}_t\le \hat{Y}_t+1$. Consider
\begin{align}
\E\left[\int_0^T (\log(\hat{Y}_t))^+\,dY_t\right]&=\E\left[\int_0^T \log(\hat{Y}^{1+}_t)\,dY_t \right]\nn
&\overset{(a)}{=}\E\left[\int_0^T \log(\hat{Y}^{1+}_t)\Gamma_t\,dt \right]\nn
&\overset{(b)}{\le} \E\left[\int_0^T \phi(\Gamma_t) - \Gamma_t+ \hat{Y}^{1+}_t\,dt\right]\nn
&\overset{(c)}{=}\E\left[\int_0^T \phi(\Gamma_t)\,dt\right] -\E\left[\int_0^T\Gamma_t\,dt\right]+\E\left[\int_0^T\hat{Y}^{1+}_t\,dt\right]\nn
&<\infty,
 \label{eq:RD_conv3}
\end{align}
where, for (a), we have used the facts that $(\hat{Y}^{1+}_t:\tin)$ is $(\F_t:\tin)$-predictable, $\log(\hat{Y}^{1+}_t)$ is non-negative, and $\Gamma_0^T$ is the $(\F_t:\tin)$-intensity of $Y_0^T$,\\*
 for (b), we note that $\hat{Y}_t^{1+}$ and $\Gamma_t$ are $P$-a.s finite, and then use the inequality in~(\ref{eq:ineq}), and \\*
for (c), we have used the facts that $\E\left[\int_0^T \phi(\Gamma_t)\,dt\right]<\infty$ (via Theorem~\ref{thm:MI}), $\E\left[\int_0^T\Gamma_t\,dt\right]<\infty$, and $\E\left[\int_0^T\hat{Y}^{1+}_t\,dt\right]\le \E\left[\int_0^T \hat{Y}_t+1\,dt\right]<\infty$.\\*
Hence we can write
\begin{align}
\E\left[\int_0^T \log(\hat{Y}_t)\,dY_t \right]
&=\E\left[\int_0^T (\log(\hat{Y}_t))^+\,dY_t\right]-\E\left[\int_0^T (\log(\hat{Y}_t))^-\,dY_t\right]\nn
&=\E\left[\int_0^T (\log(\hat{Y}_t))^+\Gamma_t\,dt \right]-\E\left[\int_0^T (\log(\hat{Y}_t))^-\Gamma_t\,dt \right]\nn
&=\E\left[\int_0^T \log(\hat{Y}_t)\Gamma_t\,dt \right].
\label{eq:rd-conv-eq}
\end{align}
\end{IEEEproof}

\begin{IEEEproof}[Proof of Lemma~\ref{le:delta_scaling_cons}]
The first part of the lemma follows directly from L'H\^{o}pital's rule. For the second part
\begin{align*}
\lim_{\Delta\to 0}\E[\bar{d}(\hat{\bar{Y}},\bar{Y})]&=\lim_{\Delta\to 0}\sum_{k=1}^4 \hat{\bar{Y}}(k)\exp(-\lambda\Delta)\alpha_k+\left(\hat{\bar{Y}}(k)-\frac{\log(\hat{\bar{Y}}(k))}{\Delta} \right)(1-\exp(-\lambda\Delta)\beta_k\\
&=\sum_{k=1}^4 \hat{\bar{Y}}(k)\alpha_k-\lambda\log(\hat{\bar{Y}}(k))\beta_k\\
&=\sum_{k=1}^4 \alpha_k \left(\hat{\bar{Y}}(k)-\frac{\lambda \beta_k}{\alpha_k}\log(\hat{\bar{Y}}(k))\right)\1\{\alpha_k>0\}\\
&\overset{(a)}{\le} \sum_{k=1}^4 \alpha_k \left( \Psi_\A\left(\frac{\lambda\beta_k}{\alpha_k}\right)+\frac{\epsilon}{4}\right)\\
&= {D} + \frac{\epsilon}{4},
\end{align*}
where for (a), we have used the definition in~(\ref{eq:CFC_Psi}).\\*
\end{IEEEproof}

\begin{IEEEproof}[Proof of Lemma~\ref{le:delta_scaling}]
The first limit can be evaluated using  L'H\^{o}pital's rule.
To compute the second limit, consider
\begin{align*}
\lim_{\Delta\to 0}{P(\bar{U}^{(i)}=k|\bar{Y}=1)}&=\lim_{\Delta\to 0}\sum_{l=0}^1{P(\bar{U}^{(i)}=k,\bar{Y}^{(i)}=l|\bar{Y}=1)}\\
&=\lim_{\Delta\to 0}\sum_{l=0}^1{P(\bar{Y}^{(i)}=l|\bar{Y}=1)}(\bar{U}^{(i)}=k|\bar{Y}^{(i)}=l)\\
&=p^{(i)}\alpha^{(i)}_{k}+(1-p^{(i)}) \beta^{(i)}_{k} \\
&={\gamma^{(i)}_k}.
\end{align*}
Then we have 
\begin{align*}
    \lim_{\Delta\to 0}P(\bar{U}_1^{(1)}=k_1,\bar{U}_1^{(2)}=k_2|\bar{Y}=1)) & =\lim_{\Delta\to 0}P(\bar{U}_1^{(1)}=k_1|\bar{Y}=1)P(\bar{U}_1^{(2)}=k_2|\bar{Y}=1) \\
    & =\gamma^{(1)}_{k_1}\gamma^{(2)}_{k_2}.
\end{align*}
Recalling that $\alpha^{(i)}_{k}=0$ implies  $\beta^{(i)}_{k}=\gamma^{(i)}_{k}=0$, we have
\begin{align*}
\lim_{\Delta\to 0}\frac{\E[\log(\hat{\bar{Y}})\1\{\bar{Y}=1\}]}{\Delta}&=\lim_{\Delta\to 0}\frac{P(\bar{Y}=1)}{\Delta}\lim_{\Delta\to 0}\E[\log(\hat{\bar{Y}})|\bar{Y}=1]\\
&=\lambda\sum_{k_1,k_2}\lim_{\Delta\to 0}P(\bar{U}=k_1,\bar{U}_2=k_2|\bar{Y}=1))\log\left(\hat{\bar{Y}}(k_1,k_2)\right)\\
&=\lambda\sum_{k_1,k_2}\gamma^{(1)}_{k_1}\gamma^{(2)}_{k_2}\log\left(\lambda\frac{\gamma^{(1)}_{k_1}\gamma^{(2)}_{k_2}}{\alpha^{(1)}_{k_1}\alpha^{(2)}_{k_2}}\right)\1\{\gamma^{(1)}_{k_1}\gamma^{(2)}_{k_2}>0\}\\
&=\lambda\sum_{k_1=1}^4\gamma^{(1)}_{k_1}\log\left(\frac{\gamma^{(1)}_{k_1}}{\alpha^{(1)}_{k_1}}\right)+\lambda\sum_{k_2=1}^4\gamma^{(1)}_{k_2}\log\left(\frac{\gamma^{(2)}_{k_2}}{\alpha^{(2)}_{k_2}}\right)+\phi(\lambda).
\end{align*}
Now to compute $ \lim_{\Delta\to 0}\E[\hat{\bar{Y}}]$, we first calculate
\begin{align*}
\lim_{\Delta\to 0}P(\bar{U}_1^{(1)}=k_1,\bar{U}_1^{(2)}=k_2)&=\lim_{\Delta\to 0}P(\bar{U}_1^{(1)}=k_1,\bar{U}_1^{(2)}=k_2|\bar{Y}=0)P(\bar{Y}=0)\\
&\quad+\lim_{\Delta\to 0}P(\bar{U}_1^{(1)}=k_1,\bar{U}_1^{(2)}=k_2|\bar{Y}=1)P(\bar{Y}=1)\\
&=\lim_{\Delta\to 0}P(\bar{U}_1^{(1)}=k_1,\bar{U}_1^{(2)}=k_2|\bar{Y}=0)\\
&=\lim_{\Delta\to 0}P(\bar{U}_1^{(1)}=k_1|\bar{Y}=0)P(\bar{U}_1^{(2)}=k_2|\bar{Y}=0)\\
&=\alpha^{(1)}_{k_1}\alpha^{(2)}_{k_2}.
\end{align*}
This gives
\begin{align*}
\lim_{\Delta\to 0}\E[\hat{\bar{Y}}]
&=\sum_{k_1,k_2}\lim_{\Delta\to 0}P(\bar{U}=k_1,\bar{U}_2=k_2)\hat{\bar{Y}}(k_1,k_2)\\
&=\lambda\sum_{k_1,k_2}\alpha^{(1)}_{k_1}\alpha^{(2)}_{k_2}\frac{\gamma^{(1)}_{k_1}\gamma^{(2)}_{k_2}}{\alpha^{(1)}_{k_1}\alpha^{(2)}_{k_2}}\1\{\alpha^{(1)}_{k_1}\alpha^{(2)}_{k_2}>0\}\\
&=\lambda.
\end{align*}
Thus
\begin{align*}
\lim_{\Delta\to 0}\E[\bar{d}(\hat{\bar{Y}},\bar{Y})]&=\lambda-\phi(\lambda)-\lambda\left( \sum_{k_1=1}^4\gamma^{(1)}_{k_1}\log\left(\frac{\gamma^{(1)}_{k_1}}{\alpha^{(1)}_{k_1}}\right)+\sum_{k_2=1}^4\gamma^{(2)}_{k_2}\log\left(\frac{\gamma^{(2)}_{k_2}}{\alpha^{(2)}_{k_2}}\right)\right)\\
&={D}.
\end{align*}
\end{IEEEproof}

\begin{IEEEproof}[Proof of Theorem~\ref{thm:Poisson_CEO}]
\subsection*{Achievability:}
Let 
\begin{gather*}
{R}^{(1)} \de \left((1-p^{(1)})\lambda+\mu^{(1)}\right) \sum_{k=1}^4 \beta^{(1)}_{k}\log\left(\frac{\beta^{(1)}_{k}}{\alpha^{(1)}_{k}}\right)\\
{R}^{(2)} \de \left((1-p^{(2)})\lambda+\mu^{(2)}\right)\sum_{k=1}^4 \beta^{(2)}_{k}\log\left(\frac{\beta^{(2)}_{k}}{\alpha^{(2)}_{k}}\right)\\
{D} \de \lambda-\phi(\lambda)-\lambda\cdot\left(\sum_{k=1}^4 \gamma^{(1)}_{k}\log\left(\frac{\gamma^{(1)}_{k}}{\alpha^{(1)}_{k}}\right)+\gamma^{(2)}_{k}\log\left(\frac{\gamma^{(2)}_{k}}{\alpha^{(2)}_{k}}\right)\right).
\end{gather*}
We will show achievability using a $(T,{R}^{(1)}+\epsilon, {R}^{(2)}+\epsilon,{D}+\epsilon)$ code without feedforward.  We will use discretization and results from the rate-distortion theory for discrete memoryless sources (DMS). 

First consider the case when for each $\iin$, at least one of the following conditions is satisfied 
\begin{enumerate}[label=C.\arabic*]
\item $\beta^{(i)}_{k}>0$ for all $k$,
\item $p^{(i)}>0$.
\end{enumerate}
Fix $\Delta>0$, and let $T\de n\Delta$ for an integer $n$.  For each $\iin$, define a binary valued discrete time process $(\bar{Y}^{(i)}_{j}:j\in\{1,\dots,n\})$ as follows. If there are one or more arrivals in the interval $((j-1)\Delta, j\Delta]$ of the process $Y_{0}^{(i), T}$, then set $\bar{Y}^{(i)}_{j}$ to $1$, otherwise set it equal to zero. Since $Y_{0}^{(i), T}$ is a Poisson process with rate $\lambda^{(i)}\de (1-p^{(i)})\lambda+\mu^{(i)}$, the components of $(\bar{Y}^{(i)}_{j}:j\in\{1,\dots,n\})$ are independent and identically distributed with $P(\bar{Y}^{(i)}=1)=1-\exp(-\lambda^{(i)}\Delta)$. Similarly, if $(\bar{Y}_{j}:j\in\{1,\dots,n\})$ denotes the discretized process $Y_0^T$, then we have
\begin{align*}
P\left(\bar{Y}^{(i)}_{j}:j\in\{1,\dots,n\}\middle|\bar{Y}_{j}:j\in\{1,\dots,n\}\right)=\prod_{j=1}^nP(\bar{Y}^{(i)}_{j}|\bar{Y}_{j})
\end{align*} 
due to the memoryless property of Poisson processes and independent thinning.
Consider  the following ``test"-channel for $\kin$,
\begin{align}
\label{EQ:test-ch1}
P(\bar{U}^{(i)}=k|\bar{Y}^{(i)}=1)=\beta^{(i)}_{k}, \\
\label{EQ:test-ch2}
P(\bar{U}^{(i)}=k|\bar{Y}^{(i)}=0)=\alpha^{(i)}_{k}.
\end{align}
Define the discretized distortion function
\begin{align}
\bar{d}(\hat{\bar{y}},\bar{y})\de \hat{\bar{y}}-\frac{\log(\hat{\bar{y}})}{\Delta}\1\{\bar{y}=1\} \quad \hat{\bar{y}}\ge 0, \bar{y}\in\{0,1\}.
\label{eq:disc_dist}
\end{align}
The reconstruction $ \hat{\bar{Y}}$ is taken as
\begin{align*}
\hat{\bar{Y}}(\bar{U}^{(1)},\bar{U}^{(2)})=\lambda \hat{\bar{Y}}^{(1)}(\bar{U}^{(1)})\hat{\bar{Y}}^{(2)}(\bar{U}^{(2)}),
\end{align*}
where
\begin{align*}
\hat{\bar{Y}}^{(i)}(k)=\begin{cases}
\frac{\gamma^{(i)}_{k}}{\alpha^{(i)}_{k}} &\text{if}\,\alpha^{(i)}_{k}>0,\\
1 &\text{otherwise}.
\end{cases}
\end{align*}
We note that since $\gamma^{(i)}_{k}=p^{(i)}\alpha^{(i)}_{k}+(1-p^{(i)})\beta^{(i)}_{k}$, and  at least one of C.1-C.2 is satisfied, $\hat{\bar{Y}}^{(i)}(k)>0$, and hence $\hat{\bar{Y}}>0$. 
Thus  the distortion function $\bar{d}(\hat{\bar{Y}},\bar{Y})$ in~(\ref{eq:disc_dist}) is bounded. 
Let
\begin{align}
\kappa\de\max_{k_1,k_2}\left|\log\left(\hat{\bar{Y}}(k_1,k_2)\right)\right|.
\label{eq:kappa_c}
\end{align}
Due to the Berger-Tung inner bound~\cite[Theorem 12.1, p. 295]{Gamal}, for a given $\Delta>0$, $\bar{\epsilon}>0$, and all sufficiently large $n$, there exists encoders $\bar{f}^{(1)}$ and $\bar{f}^{(2)}$, and a decoder $\bar{g}$ such that for $\iin$
\begin{gather*}
\bar{f}^{(i)}:(\bar{Y}^{(i)}_{j}:j\in\{1,\dots,n\})\to \{1,\dots,L^{(i)}\}\\
\bar{g}:\{1,\dots,L^{(1)}\}\times \{1,\dots,L^{(2)}\}\to(\hat{\bar{Y}}_{j}:j\in\{1,\dots,n\}),
\end{gather*}
satisfying 
\begin{align}
\frac{1}{n}\log(L^{(i)})\le I(\bar{U}^{(i)};\bar{Y}^{(i)})+\bar{\epsilon},\label{eq:dis_code1}\\
\E\left[\frac{1}{n}\sum_{j=1}^n\bar{d}(\hat{\bar{Y}}_j,\bar{Y_j})\right] \le \E[\bar{d}(\hat{\bar{Y}},\bar{Y})]+\bar{\epsilon}.
\label{eq:dis_code2}
\end{align}
It is noteworthy that the Berger-Tung inner bound has a conditioning term in the mutual-information expression, which in general is a stronger bound than that presented here. However, in our setting we can drop this conditioning as explained  in Remark~\ref{rem:BT_VNC} in the main paper.

Given the above setup, each encoder $f^{(i)}$ upon observing $Y_0^{(i),T}$ obtains  the binary valued discrete-time process $(\bar{Y}^{(i)}_{j}:j\in\{1,\dots,n\})$, and sends $M^{(i)}=\bar{f}^{(i)}(\bar{Y}^{(i)}_{j}:j\in\{1,\dots,n\})$ to the decoder. The decoder outputs the reconstruction $\hat{Y}_0^T$ as
\begin{align*}
\hat{Y}_t\de \sum_{j=1}^n\hat{\bar{Y}}_j\1\left\{t\in((j-1)\Delta, j\Delta]\right\} \quad \tin.
\end{align*}
Let $\bar{\bar{Y}}_j$ denote the actual number of arrivals of $Y_0^T$ in an interval $((j-1)\Delta,j\Delta]$. Then $\bar{d}$ is related to  the original distortion function via the above reconstruction as follows:
\begin{align*}
\frac{1}{T}d(\hat{Y}_0^T;Y_0^T)&=\frac{1}{T}\int_{0}^T\hat{Y}_t\,dt-\frac{1}{T}\int_{0}^T\log(\hat{Y}_t)\,dY_t\\
&=\frac{1}{n}\sum_{j=1}^{n}\hat{\bar{Y}}_j-\frac{1}{T}\sum_{j=1}^{n}\log(\hat{\bar{Y}}_j)\bar{\bar{Y}}_j\\
&=\frac{1}{n}\sum_{j=1}^{n}\hat{\bar{Y}}_j-\frac{1}{n\Delta}\sum_{j=1}^{n}\log(\hat{\bar{Y}}_j)\bar{Y}_j-\frac{1}{T}\sum_{j=1}^{n}\log(\hat{\bar{Y}}_j)(\bar{\bar{Y}}_j-1)\1\{\bar{\bar{Y}}_j>1\}.
\\&= \frac{1}{n}\sum_{j=1}^{n} \bar{d}(\hat{\bar{Y}}_j,\bar{Y}_j)-\frac{1}{T}\sum_{j=1}^{n}\log(\hat{\bar{Y}}_j)(\bar{\bar{Y}}_j-1)\1\{\bar{\bar{Y}}_j>1\}\\
&\overset{(a)}{\le} \frac{1}{n}\sum_{j=1}^{n} \bar{d}(\hat{\bar{Y}}_j,\bar{Y}_j)+\frac{\kappa}{T}\sum_{j=1}^{n}(\bar{\bar{Y}}_j-1)\1\{\bar{\bar{Y}}_j>1\}\\
&\le \frac{1}{n}\sum_{j=1}^{n} \bar{d}(\hat{\bar{Y}}_j,\bar{Y}_j)+\frac{\kappa}{T}\sum_{j=1}^{n}\bar{\bar{Y}}_j\1\{\bar{\bar{Y}}_j>1\},
\end{align*}
where for (a), we have used the definition of $\kappa$ in~(\ref{eq:kappa_c}).\\*
Hence taking the expectation, we get 
\begin{align}
\E\left[\frac{1}{T}d(\hat{Y}_0^T;Y_0^T)\right]&
\le \E\left[\frac{1}{n}\sum_{j=1}^{n} \bar{d}(\hat{\bar{Y}}_j,\bar{Y}_j)\right]+\kappa\E\left[\frac{1}{T}\sum_{j=1}^{n}\bar{\bar{Y}}_j\1\{\bar{\bar{Y}}_j>1\}\right]\nn
&\overset{(a)}{\le}\E[\bar{d}(\hat{\bar{Y}},\bar{Y})]+\kappa\E\left[\frac{1}{T}\sum_{j=1}^{n}\bar{\bar{Y}}_j\1\{\bar{\bar{Y}}_j>1\}\right]+\bar{\epsilon}\nn
&\overset{(b)}{=}\E[\bar{d}(\hat{\bar{Y}},\bar{Y})]+\kappa(\lambda-\lambda\exp(-\lambda\Delta))+\bar{\epsilon}\nn
&\overset{(c)}{\le} \E[\bar{d}(\hat{\bar{Y}},\bar{Y})]+\kappa\lambda^2\Delta+\bar{\epsilon},
\label{eq:ceo_ach_dbound1}
\end{align}
where, for (a), we have used~(\ref{eq:dis_code2}),\\*
for (b) we note that $\E[\bar{\bar{Y}}_j\1\{\bar{\bar{Y}}_j>1\}]=\lambda\Delta-\lambda\Delta\exp(-\lambda\Delta)$, and \\*
for (c), we have used the inequality $1-u\le \exp(-u)$. \\*
Moreover using~(\ref{eq:dis_code1}), for $\iin$ 
\begin{align}
\frac{1}{T}\log(L^{(i)})=\frac{1}{n\Delta}\log(L^{(i)})\le \frac{I(\bar{U}^{(i)};\bar{Y}^{(i)})}{\Delta}+\frac{\bar{\epsilon}}{\Delta}.
\label{eq:code-size}
\end{align}
The scaling of the mutual information $I(\bar{U}^{(i)};\bar{Y}^{(i)})$ and the distortion function $\bar{d}(\hat{\bar{Y}},\bar{Y})$ with respect to $\Delta$ is given by the following lemma.
\begin{Lemma}
For $\iin$ 
\begin{align*}
    \lim_{\Delta\to 0}\frac{I(\bar{U}^{(i)};\bar{Y}^{(i)})}{\Delta} & ={R}^{(i)},\\
    \lim_{\Delta\to 0}\E[\bar{d}(\hat{\bar{Y}},\bar{Y})] & ={D}.
\end{align*}
\label{le:delta_scaling}
\end{Lemma}
\begin{IEEEproof}
Please see the supplementary material.
\end{IEEEproof}

Now given the rate-distortion vector $({R}^{(1)}, {R}^{(2)},{D})$ and $\epsilon>0$, first choose $\Delta$ sufficiently small so that 
\begin{align*}
\frac{I(\bar{U}^{(i)};\bar{Y}^{(i)})}{\Delta}&\le{R}^{(i)}+\frac{\epsilon}{4},\\
\E[\bar{d}(\hat{\bar{Y}},\bar{Y})]&\le{D}+\frac{\epsilon}{4},\\
\kappa\lambda^2\Delta&\le\epsilon/4.
\end{align*}
Then let $\bar{\epsilon}=\Delta\epsilon/4$, and choose a sufficiently large $n$ so that~(\ref{eq:dis_code1}) and~(\ref{eq:dis_code2}) are satisfied. From~(\ref{eq:ceo_ach_dbound1}) and (\ref{eq:code-size}) we conclude that a sequence of $(T_n,{R}^{(1)}+\epsilon,{R}^{(2)}+\epsilon, {D}+\epsilon)$ code exists with $T_n=n\Delta$ when at least one of the conditions C.1 or C.2 is satisfied.

Now consider the case when $p^{(i)}=0$  some $\iin$, and for that $i$, $\beta^{(i)}_{k}=0$ for some $k$'s.  Say $p^{(1)}=0$ and $p^{(2)}>0$. This gives us $\gamma^{(1)}_k=\beta^{(1)}_k$ for $\kin$. Then we need to show that the rate-distortion vector
\begin{gather}
{R}^{(1)} = \left(\lambda+\mu^{(1)}\right) \sum_{k=1}^4 \beta^{(1)}_{k}\log\left(\frac{\beta^{(1)}_{k}}{\alpha^{(1)}_{k}}\right)\nn
{R}^{(2)} = \left((1-p^{(2)})\lambda+\mu^{(2)}\right)\sum_{k=1}^4 \beta^{(2)}_{k}\log\left(\frac{\beta^{(2)}_{k}}{\alpha^{(2)}_{k}}\right)\nn
{D} = \lambda-\phi(\lambda)-\lambda\left(\sum_{k=1}^4 \beta^{(1)}_{k}\log\left(\frac{\beta^{(1)}_{k}}{\alpha^{(1)}_{k}}\right)+\sum_{k=1}^4\gamma^{(2)}_{k}\log\left(\frac{\gamma^{(2)}_{k}}{\alpha^{(2)}_{k}}\right)\right)
\label{eq:CEO_ach_rd_vec}
\end{gather}
is achievable.
Let $[\hat{\beta}^{(1)}_k]_{k=1}^4=[1/4,1/4,1/4]$ and $[\hat{\alpha}^{(1)}_k]_{k=1}^4=[1/4,1/4,1/4-\nu,1/4+\nu]$ for some $\nu\in[0,1/3)$. Then the term 
$$
\sum_{k=1}^4 \hat{\beta}^{(1)}_{k}\log\left(\frac{\hat{\beta}^{(1)}_{k}}{\hat{\alpha}^{(1)}_{k}}\right)
$$
is continuous in $\nu$ and goes from zero to infinity as $\nu$ is increased from zero to $1/4$, hence there exists some $\hat{\nu}\in[0,1/4)$ such that with $[\hat{\alpha}^{(1)}_k]_{k=1}^4=[1/4,1/4,1/4-\hat{\nu},1/4+\hat{\nu}]$,
\begin{align}
\sum_{k=1}^4 \hat{\beta}^{(1)}_{k}\log\left(\frac{\hat{\beta}^{(1)}_{k}}{\hat{\alpha}^{(1)}_{k}}\right)=\sum_{k=1}^4 \beta^{(1)}_{k}\log\left(\frac{\beta^{(1)}_{k}}{\alpha^{(1)}_{k}}\right).
\label{eq:alpha_beta}
\end{align}
We note that this $[\hat{\beta}^{(1)}_k]_{k=1}^4$ satisfies condition C.1. Hence the rate-distortion vector in~(\ref{eq:CEO_ach_rd_vec}) is achievable by  using $[\hat{\alpha}^{(1)}_k]_{k=1}^4$  that satisfies~(\ref{eq:alpha_beta}). The case when $p^{(2)}=0$ or both $p^{(1)}=p^{(2)}=0$ can be handled similarly. 
\subsection*{Converse:}
We will prove the converse when feedforward is present.
For the given $(T,R^{(1)}+\epsilon,R^{(2)}+\epsilon,D+\epsilon)$ code with feedforward, let $M^{(1)}$ and $M^{(2)}$ denote the first and second encoder's output respectively.
We essentially repeat the steps in the converse proof of Theorem~\ref{thm:R-D} to show that 
\begin{align*}
\frac{1}{T}I(M^{(1)},M^{(2)};Y_0^T)+D \ge \lambda-\phi(\lambda)-\epsilon.
\end{align*}
Since $I(M^{(1)},M^{(2)};Y_0^T)<\infty$, we conclude from Theorem~\ref{thm:MI} that there exists 
a process $\Gamma_0^T$ such that $\Gamma_0^T$ is the $(\F_t=\sigma(M^{(1)},M^{(2)},Y_0^t):\tin)$ intensity of $Y_0^T$ and 
\begin{align}
I(M^{(1)},M^{(2)};Y_0^T)=\E\left[\int_0^T \phi(\Gamma_t)\,dt\right]-T\phi(\lambda),
\label{eq:mu_inf}
\end{align}
Let $\hat{Y}_0^T$ denote the decoder's output. The distortion constraint $D$ satisfies
\begin{align}
D\ge \frac{1}{T}\E\left[d(\hat{Y}_0^T,Y_0^T) \right]-\epsilon&=\frac{1}{T}\E\left[\int_0^T \hat{Y}_t\,dt-\log(\hat{Y}_t)\,dY_t\right]-\epsilon\nn
&=\frac{1}{T}\E\left[\int_0^T \hat{Y}_t-\log(\hat{Y}_t)\Gamma_t\,dt\right]-\epsilon,
\label{eq:CEO-conv-d}
\end{align}
where for the last equality we have used Lemma~\ref{Le:hat_Yt}.
Once again using the inequality $u\log(v)\le \phi(u)-u+v, \quad 0\le u,v < \infty$, and noting that the individual terms have finite expectations,
\begin{align}
\E\left[\int_0^T \log(\hat{Y}_t)\Gamma_t\,dt \right]&\le \E\left[\int_0^T \phi(\Gamma_t) - \Gamma_t+ \hat{Y}_t\,dt\right]\nn
&=\E\left[\int_0^T \phi(\Gamma_t)\,dt\right] - \E\left[\int_0^T\Gamma_t\,dt\right]+ \E\left[\int_0^T\hat{Y}_t\,dt\right]
 \label{eq:CEO_conv3}.
\end{align}

Combining these inequalities, we obtain
\begin{align}
\frac{1}{T}I(M^{(1)},M^{(2)};Y_0^T)+D&\ge\frac{1}{T}\E\left[\int_0^T \phi(\Gamma_t)\,dt\right]-\phi(\lambda)\nn
&\qquad+\frac{1}{T}\E\left[\int_0^T \hat{Y}_t\,dt\right]-\frac{1}{T}\E\left[\int_0^T\log(\hat{Y}_t)\,dY_t\right]-\epsilon\nn
&\overset{(a)}{\ge}\frac{1}{T}\E\left[\int_0^T\Gamma_t\,dt\right]-\phi(\lambda)-\epsilon\nn
&\overset{(b)}{=}\lambda-\phi(\lambda)-\epsilon,
\label{eq:CEO_sum_mi1}
\end{align}
where, for (a) we have used~(\ref{eq:CEO-conv-d}) and (\ref{eq:CEO_conv3}) and \\* 
for (b) we use the fact that $\E\left[\int_0^T\Gamma_t\,dt\right]=\E\left[\int_0^T\,dY_t\right]=\lambda T$.

We can upper bound the term $I(M^{(1)},M^{(2)};Y_0^T)$ as
\begin{align}
I(M^{(1)},M^{(2)};Y_0^T)&\overset{(a)}{=}H(M^{(1)},M^{(2)})-\E\left[H(M^{(1)},M^{(2)}|Y_0^T)\right]\nn
&\overset{(b)}{=}H(M^{(1)},M^{(2)})-\E\left[H(M^{(1)}|Y_0^T)\right]-\E\left[H(M^{(2)}|Y_0^T)\right]\nn
&\overset{}{\le}H(M^{(1)})+H(M^{(2)})-\E\left[H(M^{(1)}|Y_0^T)\right]-\E\left[H(M^{(2)}|Y_0^T)\right]\nn
&=I(M^{(1)};Y_0^T)+I(M^{(2)};Y_0^T),
\label{eq:CEO_sum_mi2}
\end{align}
where, for (a) we have used Lemma~\ref{Le:Wyner} and \\*
for (b), we used the Markov chain $M^{(1)}  \leftrightarrows {Y}_{0}^T \leftrightarrows M^{(2)}$.\\*
Combining (\ref{eq:CEO_sum_mi1}) and (\ref{eq:CEO_sum_mi2}) we get
\begin{align}
D\ge \lambda-\phi(\lambda)-\frac{1}{T}I(M^{(1)};Y_0^T)-\frac{1}{T}I(M^{(2)};Y_0^T)-\epsilon.
\label{eq:CEO_conv_D_ub}
\end{align}
For $\iin$, using Lemma~\ref{Le:Wyner}
\begin{align}
\frac{1}{T}I(M^{(i)};{Y}^{(i),T}_0)=\frac{1}{T}H(M^{(i)}) \le \frac{1}{T}\log\left(\lceil \exp((R^{(i)}+\epsilon)T) \rceil \right) \le R^{(i)} + \epsilon+\frac{1}{T}.
\label{eq:CEO_conv_R_ub}
\end{align}
\begin{figure}
\begin{center}
\includegraphics[scale=0.5]{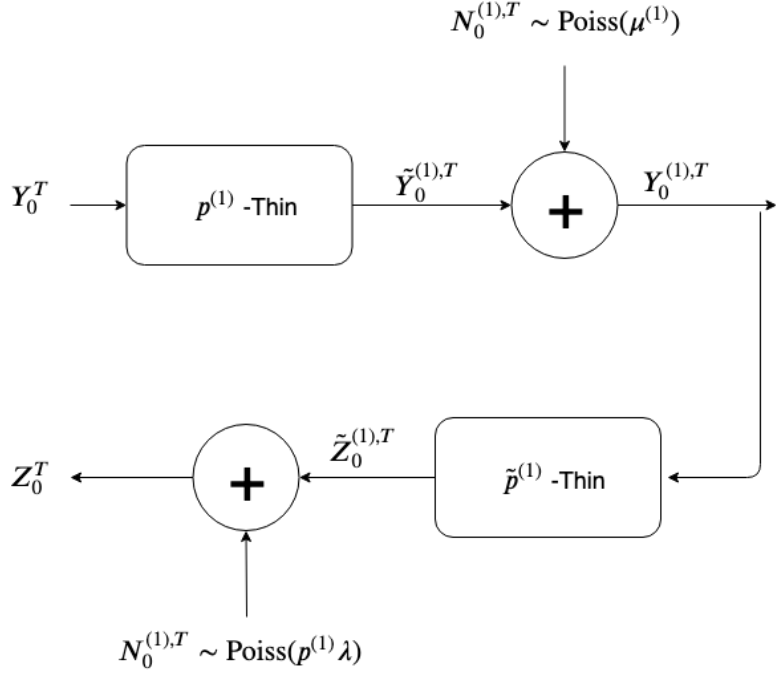}
\caption{Thinning and superposition operations defined in the proof for the first encoder. Note that the joint distribution of $(Y_{0}^{(1),T},\tilde{Y}_0^{(1),T},Y_0^T)$ is same as that of $(Y_{0}^{(1),T},\tilde{Z}_0^{(1),T},Z_0^T)$.} 
\label{Fig:CEO1}
\end{center}
\end{figure}
We will first consider the case when $p^{(i)}<1$ for $\iin$. We shall proceed by defining certain auxiliary processes (see Figure~\ref{Fig:CEO1}).
Let $\tilde{Z}_0^{(i),T}$ be obtained from $\tilde{p}^{(i)}$-thinning of  $Y_{0}^{(i), T}$, where
$$
\tilde{p}^{(i)}=\frac{\mu^{(i)}}{((1-p^{(i)})\lambda+\mu^{(i)})}.
$$
Then using Lemma~\ref{le:Poiss_Split} we can write
$$
Y_{t}^{(i)}=\tilde{Z}_{t}^{(i)} + \hat{Z}_{t}^{(i)} \quad \tin,
$$
where $\tilde{Z}_{0}^{(i),T}$ and $\hat{Z}_{0}^{(i),T}$ are independent Poisson processes with rates $(1-p^{(i)})\lambda$ and $\mu^{(i)}$ respectively.
Whereas, by definition
$$
Y_{t}^{(i)}=\tilde{Y}_{t}^{(i)} + {N}_{t}^{(i)} \quad \tin,
$$
where $\tilde{Y}_{t}^{(i)}$ and $N_{t}^{(i)}$ are independent Poisson processes with rates $(1-p^{(i)})\lambda$ and $\mu^{(i)}$ respectively.
Hence we conclude that the joint distribution of $(Y_{0}^{(i),T},\tilde{Z}_{0}^{(i),T})$ is identical to the joint distribution of  $(Y_{0}^{(i),T},\tilde{Y}_{0}^{(i),T})$. 
Let $Z_0^{(i),T}$ be obtained  by adding an independent Poisson process $\hat{N}_{0}^{(i),T}$ with rate $p^{(i)}\lambda$ to   $\tilde{Z}^{(i),T}_0$, 
$$
Z_t^{(i)}=\tilde{Z}_{t}^{(i)}+\hat{N}_{t}^{(i)}\quad \tin.
$$
Also using Lemma~\ref{le:Poiss_Split} we have
$$
Y_t=\tilde{Y}^{(i)}_t+\tilde{\tilde{Y}}^{(i)}_t\quad \tin,
$$
where $\tilde{Y}_0^{(i),T}$ and $\tilde{\tilde{Y}}^{(i),T}_0$ are independent Poisson processes with rates $(1-p^{(i)})\lambda$ and $p^{(i)}\lambda$. 
Hence the joint distribution of $(Z_0^{(i),T}, \tilde{Z}^{(i),T}_0)$ and $(Y_0^T, \tilde{Y}^{(i),T}_0)$ are identical.
Moreover, $M^{(i)}\rightleftarrows Y_{0}^{(i),T} \rightleftarrows \tilde{Y}_0^{(i),T} \rightleftarrows Y_0^T$ forms a Markov chain and 
$M^{(i)}\rightleftarrows Y_{0}^{(i),T} \rightleftarrows \tilde{Z}_0^{(i),T} \rightleftarrows Z_0^{(i),T}$ forms a Markov chain.
This allows us to write
\begin{align}
I\left(M^{(i)}; \tilde{Z}_0^{(i),T}\right)&=I\left(M^{(i)}; \tilde{Y}_0^{(i),T}\right),\nn
I\left(M^{(i)}; {Z}_0^{(i),T}\right) &= I(M^{(i)};Y_0^T).
\label{eq:same_mu_inf}
\end{align}

Since $\tilde{Z}_0^{(i),T}$ is a $\frac{\mu^{(i)}}{((1-p^{(i)})\lambda+\mu^{(i)})}$-thinning of ${Y}^{(i),T}_0$, Theorem~\ref{thm:thin_int} gives
\begin{align}
I(M^{(i)};\tilde{Z}_0^{(i),T})\le \left(1-\frac{\mu^{(i)}}{(1-p^{(i)})\lambda+\mu^{(i)}}\right)I(M^{(i)};{Y}^{(i),T}_0).
\label{eq:thin_Poiss}
\end{align}
Also $Z_0^{(i),T}$ is obtained  by adding an independent Poisson process  with rate $p^{(i)}\lambda$ to   $\tilde{Z}^{(i),T}_0$, Theorem~\ref{thm:add_int} yields
\begin{align}
I(M^{(i)};\tilde{Z}_0^{(i),T})&=\E\left[\int_0^T \phi(\tilde{\Gamma}_{t}^{(i)})-\phi((1-p^{(i)})\lambda)\,dt\right], \nn
I(M^{(i)};{Z}_0^{(i),T})&\le\E\left[\int_0^T \phi(\tilde{\Gamma}_{t}^{(i)}+p^{(i)}\lambda)-\phi(\lambda)\,dt\right],
\label{eq:add_Poiss}
\end{align}
where, $\tilde{\Gamma}_0^{(i),T}$ is the $(\sigma(M^{(i)},\tilde{Z}_{0}^{(i), T}):\tin)$-intensity of $\tilde{Z}_0^{(i),T}$.
Then we can further lower bound $D$ in~(\ref{eq:CEO_conv_D_ub}) as 
\begin{align*}
D&\ge \lambda-\phi(\lambda)-\frac{1}{T}I(M^{(1)};Y_0^T)-\frac{1}{T}I(M^{(2)};Y_0^T)-\epsilon\\
&\overset{(a)}{=} \lambda-\phi(\lambda)-\frac{1}{T}I(M^{(1)};{Z}_0^{(1),T})-\frac{1}{T}I(M^{(2)};{Z}_0^{(2),T})-\epsilon\\
&\overset{(b)}{\ge} \lambda-\phi(\lambda)-\frac{1}{T}\left(\E\left[\int_0^T \phi(\tilde{\Gamma}_{t}^{(1)}+p^{(1)}\lambda)-\phi(\lambda)\,dt\right]\right) \\
    & \phantom{===} -\frac{1}{T}\left(\E\left[\int_0^T \phi(\tilde{\Gamma}_{t}^{(2)}+p^{(2)}\lambda)-\phi(\lambda)\,dt\right]\right)-\epsilon\\
&\overset{(c)}{=} \lambda+\phi(\lambda)-\E\left[\phi(\tilde{\Gamma}_{S_1}^{(1)}+p^{(1)}\lambda)\right]-\E\left[\phi(\tilde{\Gamma}^{(2)}_{S_2}+p^{(2)}\lambda)\right]-\epsilon,
\end{align*}
where for (a), we have used~(\ref{eq:same_mu_inf}),\\*
 for (b), we have used~(\ref{eq:add_Poiss}), and \\*
for (c), we define $S_1$ and $S_2$ to be uniformly distributed on $[0,T]$,  independent of all other random variables and independent of each other as well.\\*
For each $\iin$, $R^{(i)}$ in~(\ref{eq:CEO_conv_R_ub}) can be lower bounded as  
\begin{align*}
R^{(i)}&\ge \frac{1}{T}I(M^{(i)};{Y}^{(i),T}_0)-\epsilon-\frac{1}{T}\\
&\overset{(a)}{\ge}\frac{(1-p^{(i)})\lambda+\mu^{(i)}}{(1-p^{(i)})\lambda}\frac{1}{T}I(M^{(i)};\tilde{Z}_0^{(i),T})-\epsilon-\frac{1}{T}\\
&\overset{(b)}{=}\frac{(1-p^{(i)})\lambda+\mu^{(i)}}{(1-p^{(i)})\lambda}\frac{1}{T}\E\left[\int_0^T \phi(\tilde{\Gamma}_{t}^{(i)})-\phi((1-p^{(i)})\lambda)\,dt\right]-\epsilon-\frac{1}{T}\\
&\overset{(c)}{=}\frac{(1-p^{(i)})\lambda+\mu^{(i)}}{(1-p^{(i)})\lambda}\E\left[\phi(\tilde{\Gamma}_{S_i}^{(i)})-\phi((1-p^{(i)})\lambda)\right]-\epsilon-\frac{1}{T},
\end{align*}
where for (a), we have used~(\ref{eq:thin_Poiss}),\\*
for (b), we have used~(\ref{eq:add_Poiss}), and \\*
for (c), recall that  $S_1$ and $S_2$ are uniformly distributed on $[0,T]$,  independent of all other random variables and independent of each other.\\*
Now we use Carath\'{e}odory's  theorem~\cite[Theorem 17.1]{Rockafellar'97}. For each $\iin$, there exist  non-negative $[\eta^{(i)}_k]_{k=1}^4$ and $[\alpha^{(i)}_k]_{k=1}^4$, such that $\sum_{k=1}^4 \alpha^{(i)}_k =1$ and 
\begin{align*}
\E\left[\phi(\tilde{\Gamma}_{S_i}^{(i)})\right]&=\sum_{k=1}^4 \alpha^{(i)}_k\phi(\eta^{(i)}_k),\\
\E\left[\phi(\tilde{\Gamma}_{S_i}^{(i)}+p^{(i)}\lambda)\right]&=\sum_{k=1}^4 \alpha^{(i)}_k\phi(\eta^{(i)}_k+p^{(i)}\lambda),\\
\E\left[\tilde{\Gamma}_{S_i}^{(i)}\right]&=\sum_{k=1}^4 \alpha^{(i)}_k\eta^{(i)}_k=(1-p^{(i)})\lambda,
\end{align*}
where in the last line we have used the fact that since $\tilde{\Gamma}_0^{(i),T}$ is the $(\sigma(M^{(i)},\tilde{Z}_{0}^{(i), T}):\tin)$-intensity of $\tilde{Z}_0^{(i),T}$,
$\E\left[\int_0^T\tilde{\Gamma}_{t}^{(i)}\,dt\right]=\E[\tilde{Z}^{(i)}_T]=T(1-p^{(i)})\lambda$.
Hence we have
\begin{align}
R^{(i)}&\ge\frac{(1-p^{(i)})\lambda+\mu^{(i)}}{(1-p^{(i)})\lambda}\left(\sum_{k=1}^4 \alpha^{(i)}_k\phi(\eta^{(i)}_k)-\phi((1-p^{(i)})\lambda)\right)-\epsilon-\frac{1}{T},\label{eq:rdconv_r-i_1}\\
D &\ge \lambda+\phi(\lambda)-\sum_{k=1}^4 \alpha^{(1)}_k\phi(\eta^{(1)}_k+p^{(1)}\lambda)-\sum_{k=1}^4 \alpha^{(2)}_k\phi(\eta^{(2)}_k+p^{(2)}\lambda)-\epsilon. \label{eq:rdconv_d_1}
\end{align}
Now define 
$$\beta^{(i)}_k\de\frac{\alpha^{(i)}_k\eta^{(i)}_k}{(1-p^{(i)})\lambda}, \quad \gamma^{(i)}_{k}\de p^{(i)}\alpha^{(i)}_{k}+(1-p^{(i)})\beta^{(i)}_{k}.$$ 
We note that $\beta^{(i)}_k=0$ if $\alpha^{(i)}_k=0$, and $\sum_{k=1}^4\beta^{(i)}_k=1$. Substituting the above definitions in~(\ref{eq:rdconv_r-i_1})
 \begin{align}
 R^{(i)} &\ge\frac{(1-p^{(i)})\lambda+\mu^{(i)}}{(1-p^{(i)})\lambda} \left(\sum_{k=1}^4 \alpha^{(i)}_k\eta^{(i)}_k\log(\eta^{(i)}_k)-\phi((1-p^{(i)})\lambda)\right)-\epsilon-\frac{1}{T}\nn
 &=((1-p^{(i)})\lambda+\mu^{(i)})\left( \sum_{k=1}^4 \beta^{(i)}_k\log\left(\frac{\beta^{(i)}_k(1-p^{(i)})\lambda}{\alpha^{(i)}_k}\right)\1\{\alpha^{(i)}_k>0\}-\log((1-p^{(i)})\lambda)\right) \\
     & \phantom{=========} -\epsilon-\frac{1}{T}\nn
 &=((1-p^{(i)})\lambda+\mu^{(i)}) \sum_{k=1}^4 \beta^{(i)}_k\log\left(\frac{\beta^{(i)}_k}{\alpha^{(i)}_k}\right)-\epsilon-\frac{1}{T}.
\label{eq:rdconv_r-i_2} 
 \end{align}
Likewise,
\begin{align*}
\sum_{k=1}^4\alpha^{(i)}_k\phi(\eta^{(i)}_k+p^{(i)}\lambda)&=\sum_{k=1}^4\alpha^{(i)}_k\phi\left(\frac{\beta^{(i)}_k(1-p^{(i)})\lambda}{\alpha^{(i)}_k}+p^{(i)}\lambda\right)\1\{\alpha^{(i)}_k>0\}\\
&=\sum_{k=1}^4\alpha^{(i)}_k\phi\left(\frac{\gamma^{(i)}_k}{\alpha^{(i)}_k}\lambda\right)\1\{\alpha^{(i)}_k>0\}\\
&=\lambda\sum_{k=1}^4\gamma^{(i)}_k\log\left(\frac{\gamma^{(i)}_k}{\alpha^{(i)}_k}\right)+\phi(\lambda).
\end{align*}
Substituting the above in~(\ref{eq:rdconv_d_1}), we get
\begin{align}
D \ge \lambda-\phi(\lambda)-\lambda\sum_{k=1}^4\gamma^{(1)}_k\log\left(\frac{\gamma^{(1)}_k}{\alpha^{(1)}_k}\right)-\lambda\sum_{k=1}^4\gamma^{(2)}_k\log\left(\frac{\gamma^{(2)}_k}{\alpha^{(2)}_k}\right)-\epsilon.
\label{eq:rdconv_d_2}
\end{align}

If either $p^{(i)}$, say $p^{(1)}$, equals 1, then $M^{(1)}$ and $Y_0^T$ are independent so that $I(M^{(1)};Y_0^T)=0$, and  we can repeat the above steps   to show that 
\begin{align*}
 R^{(2)} &\ge  ((1-p^{(2)})\lambda+\mu^{(2)}) \sum_{k=1}^4 \beta^{(2)}_k\log\left(\frac{\beta^{(2)}}{\alpha^{(2)}_k}\right)-\epsilon-\frac{1}{T}, \\
 D &\ge \lambda-\phi(\lambda)-\lambda\sum_{k=1}^4\gamma^{(2)}_k\log\left(\frac{\gamma^{(2)}_k}{\alpha^{(2)}_k}\right)-\epsilon,
 \end{align*}
 which is the region  in (\ref{eq:rdconv_r-i_2})-(\ref{eq:rdconv_d_2})  with $\alpha^{(1)}_k=\beta^{(1)}_k=\gamma^{(1)}_k$ for $\kin$. 
 
 Since $\epsilon$ is arbitrary, taking $\epsilon\to 0$ and $T\to\infty$ gives us the rate region in the statement of the theorem. 
 
\end{IEEEproof}
\end{document}